\documentclass[a4paper,12pt]{article}

\newlength{\dinwidth}
\newlength{\dinmargin}
\begin{document}

\bigskip

\title{Branching Ratio and CP Violation of 
$B\to \pi \pi$ Decays \\
in Perturbative QCD Approach}

\author{Cai-Dian L\"u$^a$\footnote{e-mail: lucd@theo.phys.sci.hiroshima-u.ac.jp},  Kazumasa Ukai$^b$\footnote{e-mail: ukai@eken.phys.nagoya-u.ac.jp}, Mao-Zhi Yang$^a$\footnote{e-mail: yangmz@theo.phys.sci.hiroshima-u.ac.jp}\\
{\it \small $a$ Physics Department, Hiroshima University,
Higashi-Hiroshima 739-8526, Japan}\\
{\it \small $b$ Physics Department, Nagoya University, Nagoya 464-8602, Japan}}

\maketitle

\begin{picture}(0,0)
       \put(335,320){\bf hep-ph/0004213}
       \put(335,300){HUPD-9924}
       \put(335,280){DPNU-00-15}
\end{picture}

\begin{abstract}
We calculate the branching ratios and CP asymmetries for $B^0 \to \pi^+\pi^-$,
$B^+ \to \pi^+\pi^0$ and $B^0 \to \pi^0\pi^0$ decays, in a 
perturbative QCD approach. In this approach, we calculate non-factorizable 
and annihilation type contributions, in addition to the usual factorizable 
contributions. 
We found that the annihilation diagram contributions are not very small as
previous argument.
Our result is in agreement with the measured branching ratio 
of $B \to \pi^+\pi^-$ by CLEO collaboration. 
With a non-negligible contribution from annihilation diagrams and a
large strong phase, we predict a
large direct CP asymmetry in $B^0 \to \pi^+\pi^-$, and $\pi^0\pi^0$,
which can be tested by the current running B factories.
\end{abstract}

\bigskip

PACS: 13.25.Hw, 11,10.Hi, 12,38.Bx, 

\newpage

\section{Introduction}

The charmless $B$ decays arouse more and more interests recently, since 
it is a good place for study of CP violation and it is also sensitive 
to new physics \cite{sanda}.
Factorization approach (FA) is applied to hadronic $B$ decays 
and is generalized to decay modes that are classified in the spin 
of final states \cite{fac,akl1,cheng}.
FA gives predictions in terms of form factors and decay constants.
Although the predictions of branching ratios agree well with experiments
in most cases, there are still some theoretical points  unclear.
First, it relies strongly  on the form factors, which cannot be 
calculated by FA itself. 
Secondly, the generalized FA 
shows that the non-factorizable contributions are important in
a group of channels \cite{akl1,cheng}.
The reason of this large non-factorizable contribution needs more 
theoretical studies. 
Thirdly, the strong phase, which
is important for the CP violation prediction, is quite sensitive to
 the internal gluon momentum \cite{akl2}.
This gluon momentum is the sum of  momenta of two quarks, which go into 
two different mesons. 
It is difficult to define exactly in the FA approach.
To improve the theoretical predictions of the non-leptonic
$B$ decays, we try to improve the factorization approach, and explain the
size of the non-factorizable contributions in a new approach.

We shall take a specific channel $B\to \pi\pi$ as an example.
The $B\to \pi\pi$ decays are responsible for the determination of 
the angle $\phi_2$ in the unitarity triangle which have been studied in 
the factorization approach in detail \cite{fac,akl1,cheng}. 
The recent measurements of $B\to \pi^+\pi^-$ by CLEO Collaboration 
attracted much attention for these kind of decays \cite{cleo}.
The most recent theoretical study \cite{neubert} attempted to compute
the non-factorizable diagrams directly. 
But it could not also predict the transition form factors of $B\to \pi$.

In this paper, we would like to study the $B\to \pi\pi$ decays in the 
perturbative QCD approach (PQCD) \cite{7}.
In the $B\to \pi\pi$ decays, the $B$ meson is heavy, sitting at rest. 
It decays into two light mesons with large momenta. Therefore the light 
mesons are moving very fast in the rest frame of $B$ meson. 
In this case, the short distance hard process dominates the decay amplitude.
We shall demonstrate that
the soft final state interaction is not important, 
since there is not enough time for the pions to exchange soft gluons.
This makes the perturbative QCD approach applicable.
With the final pions moving very fast, there must be a hard gluon to kick 
the light spectator quark $d$ or $u$ (almost at rest) 
in the $B$ meson to form a fast moving pion 
(see Figure \ref{fig:PhysicalPicB_1}).
So the dominant diagram in this theoretical picture is that one hard gluon
 from the spectator quark connecting with the other quarks in the four
quark operator of the weak interaction.
Unlike the usual FA, where the spectator quark does 
not participate in the decay process in a major way, the hard part of 
the PQCD approach
consists of six quarks rather than four.
We thus call it six-quark operators or six-quark effective theory.
Applying the six-quark effective theory to $B$ meson exclusive
decays, we need meson 
wave functions for the hadronization of quarks into mesons.
Separating that nonperturbative dynamics from the hard one, the decay
amplitudes can be  calculated in PQCD easily.
Most of the nonperturbative dynamics are included in the meson wave 
functions, but in the correction that soft gluon straddle the six-quark
operators, there are some nonfactorizable soft gluon effects not to be 
absorbed into the meson wave functions.
Such effects can be safely neglected in the $B$ meson decays \cite{LiTseng}.

Li performed the calculation of $\bar{B}^0 \to \pi^+ \pi^-$ in
ref.\cite{pp} using the PQCD formalism, where the factorizable tree
diagrams were calculated and the branching ratios were predicted.
In another paper \cite{kroll}, Dahm, Jakob and Kroll performed a more
complete calculation, including the non-factorizable annihilation
topology and the three decay channels of $B\to \pi \pi$ decays. 
However, the predicted branching ratios are about
one order smaller than the current experiments by CLEO \cite{cleo}. 
In connection with this, Feldmann and Kroll concluded that 
perturbative contributions to the $B\to \pi$ transition form factor 
were much smaller than nonperturbative ones \cite{Feldmann:1999sm}.
As we shall show later, 
the pion wave function must be consistent with chiral symmetry relation
\begin{equation}
 -q^\mu \langle 0| \bar{u} \gamma_\mu \gamma_5 d(x) |\pi^-(q) \rangle
 = (m_u + m_d)\langle 0| \bar{u}\gamma_5 d(x) |\pi^-(q) \rangle .
\end{equation}
This introduces terms that were not considered in above calculations.
In this paper, considering the terms needed from chiral symmetry, 
we calculate the $B\to \pi$ transition form factors and also 
the non-factorizable contributions in PQCD approach.
We then show that our result for the
branching ratio $B \to \pi^+\pi^-$ agree with the measurement.
Among the new terms, it is worthwhile emphasizing the presence of 
annihilation diagrams which are ignored in FA.
We find that these diagrams can not be ignored, and furthermore 
they contribute to large final state interaction phase.

\section{The Frame Work}

The three scale 
PQCD factorization theorem has been developed for non-leptonic heavy
meson decays \cite{li}, based on the formalism by Brodsky and Lepage \cite{bl},
and Botts and Sterman \cite{bs}.
The QCD corrections to the four quark operators are usually summed by
the renormalization group equation \cite{buras}.
This has already been done to the leading logarithm and next-to-leading
order for years.
Since the $b$ quark decay scale $m_b$ is much smaller than the electroweak
scale $m_W$, the QCD corrections are non-negligible.
The third scale $1/b$ involved in the $B$ meson exclusive decays is 
usually called the factorization scale, 
with $b$ the conjugate variable of parton transverse momenta.
The dynamics below $1/b$ 
 scale is regarded as being completely 
nonperturbative, and can be parametrized into meson wave functions.
The meson wave functions are not calculable in PQCD.
But they are universal, channel independent. 
We can determine it from experiments,
and it is constrained   by QCD sum rules and Lattice QCD calculations.
Above the scale $1/b$, the physics is channel dependent. 
We can use perturbation theory to calculate channel by channel.

Besides the hard gluon exchange with the spectator quark, 
the soft gluon exchanges between quark lines  give 
out the double logarithms $\ln^2(Pb)$ from the overlap of collinear
and soft divergences, $P$ being the dominant light-cone component of a 
meson momentum.
The resummation of these double logarithms leads to a Sudakov form factor
$\exp[-s(P,b)]$, which suppresses the long distance contributions in the 
large $b$ region,
and vanishes as $b> 1/\Lambda_{QCD}$.
This form factor  is given to sum the leading order
soft gluon exchanges between the hard part and the wave functions of
mesons. So this term includes the double infrared logarithms. 
The expression of $s(Q,b)$ is concretely given in appendix 
\ref{ap:FormulasHardPart}.
Figure \ref{fig:sudakov} shows that 
$e^{-s}$ falls off quickly in the large $b$, or
long-distance, region, giving so-called Sudakov suppression.
This Sudakov factor practically makes PQCD approach applicable.
For the detailed derivation of the Sudakov form factors, see 
ref.\cite{7,8}.

With all the large logarithms resummed, the remaining finite contributions are 
absorbed into a perturbative b quark decay subamplitude $H(t)$. 
Therefore the three scale 
factorization formula is given by the typical expression,
\begin{equation}
C(t) \times H(t) \times \Phi (x) \times \exp\left[ -s(P,b) 
-2 \int _{1/b}^t \frac{ d \bar\mu}{\bar \mu} \gamma_q (\alpha_s (\bar \mu)) 
\right], \label{eq:factorization_formula}
\end{equation}
where $C(t)$ are the corresponding Wilson coefficients, 
$\Phi (x)$ are the  meson wave functions and
the variable $t$ denotes the largest mass scale of hard process $H$, 
that is, six-quark effective theory.
 The quark anomalous dimension $\gamma_q=-\alpha_s /\pi$ describes 
the evolution from scale $t$ to $1/b$. 
Since logarithm corrections have been summed by renormalization group 
equations, the above factorization formula does not depend on the 
renormalization scale $\mu$ explicitly. 

The three scale factorization theorem in
eq.(\ref{eq:factorization_formula}) is  discussed by Li {\it et al}. 
in detail \cite{li}.
Below section \ref{sc:PertCalc}, we shall give the factorization
formulae for $B\to \pi\pi$ decay amplitudes by calculating the hard part
$H(t)$, channel dependent in PQCD.
We shall also approximate $H$ there by the ${\cal O}(\alpha_s)$
expression, which makes sense if perturbative contributions indeed
dominate.

In the resummation procedures, the $B$ meson is treated as a heavy-light
system. The wave function is defined as
\begin{equation}
\Phi_B= \frac{1}{\sqrt{2N_c}} 
(\not \! p_B +m_B) \gamma_5 \phi_B (k_1,k_b),
\end{equation}
where $N_c=3$ is color's degree of freedom and 
$\phi_B (k_1,k_b)$ is the distribution function of the 
4-momenta of the light quark ($k_1$) and $b$ quark ($k_b$)
\begin{equation}
\phi_B(k_1, k_b) = \frac{1}{ p_B^2 }
\frac{1}{2\sqrt{2N_c}}
 \int\!\! \frac{d^4 y}{(2\pi)^4}
 e^{ik_1 \cdot y}
 \langle 0| {\rm T}[\bar{d}(y)\! \not \! p_B \gamma_5 
 b(0)]|B(p_B)\rangle . 
 \label{eq:BDistFunc}
\end{equation}
Note that we use the same distribution function $\phi_B (k_1,k_b)$
for the $\not \! p_B$ term and the $m_B$ term 
 from heavy quark effective theory. 
For the hard part calculations in the next section, 
we use the approximation $m_b\simeq m_B$, which is the same order
approximation neglecting higher twist of $(m_B-m_b)/m_B$.
To form a bound state of $B$ meson, the condition $k_b=p_B-k_1$ is required.
So $\phi_B$ is actually a function of $k_1$ only. 
Through out this paper, we take $p^\pm = (p^0 \pm p^3)/\sqrt{2}$, 
${\bf p}_T = (p^1, p^2)$ as the light-cone coordinates 
to write the four momentum.
We consider the $B$ meson at rest, then that momentum is 
$p_B=(m_B/\sqrt{2}) (1,1,{\bf 0}_T)$. 
The momentum of the light valence quark is written as
 ($k_1^+,k_1^-, {\bf k}_{1T}$), where the $ {\bf k}_{1T}$ 
is a small transverse 
momentum. 
It is difficult to define the function $\phi_B(k_1^+,k_1^-,{\bf k}_{1T})$.
However, in the next section, we will see
that the hard part is always independent of $k_1^+$,
if we make some approximations. 
This means that $k_1^+$ can be integrated out in
eq.(\ref{eq:BDistFunc}), the function $\phi_B(k_1^+,k_1^-,{\bf k}_{1T})$ 
can be simplified to 
\begin{eqnarray}
\phi_B (x_1, {\bf k}_{1T}) &=& p_B^-
\int d k_1^+ \phi_B (k_1^+, k_1^-, {\bf k}_{1T}) \nonumber \\
 &=& \frac{p_B^-}{ p_B^2 } \frac{1}{2\sqrt{2N_c}}
 \int\!\! \frac{d y^+ d^2 {\bf y}_T}{(2\pi)^3}
 e^{i (k_1^- y^+ - {\bf k}_{1T}\cdot{\bf y}_T)}
 \langle 0| {\rm T}[\bar{d}(y^+, 0, {\bf y}_T)\! \not \! p_B \gamma_5 
 b(0)]|B(p_B)\rangle,
\label{int}
\end{eqnarray}
where $x_1=k_1^-/p_B^-$ is the momentum fraction.
Therefore,
in the perturbative calculations, we do not need the information of all
four momentum $k_1$.
The above integration can be done only when the hard part of the subprocess
is independent of the variable $k_1^+$. 

The $\pi$ meson is treated as a light-light system. 
At the $B$ meson rest frame, pion is moving very fast. 
We define the momentum of the pion which contain 
the spectator light quark as $P_2= (m_B/\sqrt{2}) (1,0, {\bf 0}_T)$. 
The other pion which moving to the inverse direction, then has momentum
$P_3= (m_B/\sqrt{2}) (0,1, {\bf 0}_T)$. 
The light spectator quark moving with the pion (with momentum $P_2$),
has a momentum $(k_2^+, 0, {\bf k}_{2T})$. 
The momentum of the other valence quark in this pion is then
$(P_2^+ - k_2^+, 0, -{\bf k}_{2T})$. 
 If we define the momentum fraction as $x_2=k_2^+ / P_2^+$, then
the wave function of pion can be written as
\begin{equation}
\Phi_\pi= \frac{1}{\sqrt{2N_c}} \gamma_5 \left[\not \! p_\pi 
\phi_\pi (x_2,{\bf k}_{2T}) 
+ m_0 \phi'_\pi (x_2,{\bf k}_{2T}) \right],
\label{eq:wf_pion}
\end{equation}
where $\phi_\pi (x_2,{\bf k}_{2T})$ is defined in analogue to 
eq.(\ref{eq:BDistFunc}, \ref{int}) and $\phi'_\pi (x_2,{\bf k}_{2T})$
is defined by
\begin{equation}
\phi'_\pi(x_2,{\bf k}_{2T})
= \frac{P_2^+}{2\sqrt{2N_c}}
 \int\! \frac{d y^- d^2 {\bf y}_T }{(2\pi)^3}
 e^{i( x_2P_2^+ y^- - {\bf k}_{2T} \cdot{\bf y}_T)}
 \langle 0| {\rm T}[\bar{d}(0) \gamma_5 
 u(0,y^-,{\bf y}_T)]|\pi(P_2)\rangle .
\end{equation}
Note that as you shall see below, $m_0$ given as
\begin{equation}
 m_0=\frac{m_\pi^2}{m_u+m_d}
\label{eq:def_m0}
\end{equation}
in eq.(\ref{eq:wf_pion}) {\it is not} the pion mass.
Since this $m_0$ is estimated around $1\sim 2$ GeV using the quark masses
predicted from lattice simulations,
one may guess contributions of $m_0$ term cannot be neglected because of 
$m_0 \not \!\ll m_B$.
In fact, we will show this $m_0$ plays important roles to predict 
the $B\to\pi\pi$ branching
ratios in section \ref{sc:NumCalc}.

The normalization of wave functions is determined by meson's decay
constant
\begin{equation}
\langle 0| \bar d(0) \gamma_\mu \gamma_5 u (0) | \pi (p) \rangle
= i p_\mu f_\pi .
\label{eq:PCAC}
\end{equation}
Using this relation, the normalization of $\phi_\pi$ is defined as
\begin{equation}
 \int dx_2 d^2{\bf k}_{2T} \phi_\pi (x_2,{\bf k}_{2T})
= \frac{f_\pi}{2\sqrt{2N_c}} .
\label{eq:norm_pi}
\end{equation}
Moreover, from eq.(\ref{eq:PCAC}) you can readily derive
\begin{equation}
\langle 0| \bar d(0)  \gamma_5 u (0) | \pi (p) \rangle
= -i \frac{m_\pi^2}{m_u+m_d} f_\pi ,
\end{equation}
so defining $m_0$ such as eq.(\ref{eq:def_m0}), the normalization of
$\phi'_\pi$ is the same one to eq.(\ref{eq:norm_pi}).

The transverse momentum ${\bf k}_T$ is usually conveniently 
converted to the $b$ parameter by
Fourier transformation.
 The initial conditions of $\phi^{(\prime)}_i(x)$,
$i=B$, $\pi$, are of nonperturbative origin, satisfying the
normalization
\begin{equation}
\int_0^1\phi^{(\prime)}_i(x,b=0)dx=\frac{f_i}{2\sqrt{2N_c}}\;,
\label{no}
\end{equation}
with $f_i$ the meson decay constants.

\section{Perturbative Calculations}
\label{sc:PertCalc}

With the above brief discussion, the only thing left is to compute 
$H$ for each diagram. 
There are altogether 8 diagrams contributing to the $B\to \pi\pi$ decays,
which are shown in Figure \ref{fig:diagrams}.
They are the lowest order diagrams. 
In fact the diagrams without hard gluon exchange between the spectator quark
and other quarks are suppressed by the wave functions. 
The reason is that the light quark in $B$ meson is almost at rest. 
If there is no large momentum exchange with other quarks, 
it carries almost zero momentum in the fast moving $\pi$, that is the
end point of pion wave function. 
In the next section, we will see that the pion wave function at the zero 
point is always zero. 
The Sudakov form factor suppresses the large number of soft gluons 
exchange to transfer large momentum.
It is already shown that the hard gluon is really hard in the 
numerical calculations of $B\to K \pi $ \cite{kls}. 
The value of  $\alpha_s /\pi$ is peaked below $0.2$.
And in our following calculation of $B\to \pi \pi $ decays 
this is also proved.

Let's first calculate the usual factorizable diagrams (a) and (b).
The four quark operators indicated by a cross in the diagrams, are shown
in the appendix A. 
There are  two kinds of operators. Operators $O_1$, $O_2$, $O_3$,
 $O_4$, $O_9$, and $O_{10}$ are $(V-A)(V-A)$ currents,  the sum of their
amplitudes is given as
\begin{eqnarray}
F_e &=& - 16\pi C_F m_B^2
\int_{0}^{1}d x_{1}d x_{2}\,\int_{0}^{\infty} b_1d b_1 b_2d b_2\,
\phi_B(x_1,b_1)
\nonumber \\
& &\times \left\{ \left[(1+x_2)\phi_\pi(x_2,b_2)
 + (1-2x_2) \phi '_\pi(x_2,b_2) r_\pi\right] \right.
\nonumber \\
& &\times \alpha_{s}(t_e^1)
h_e (x_1,x_2,b_1,b_2) \exp[-S_{B}(t_e^1)-S_\pi^1(t_e^1)]
+2 r_\pi \phi '_\pi(x_2,b_2)
\nonumber \\
& &\times \left .\alpha_{s}(t_e^2)
 h_e (x_2,x_1,b_2,b_1) \exp[-S_{B}(t_e^2)-S_\pi^1(t_e^2)] \right \} ,
\label{a}
\end{eqnarray}
where $r_\pi =m_0 / m_B =m_\pi^2 / [m_B (m_u+m_d) ]$. 
$C_F=4/3$ is a color factor.
The function $h_e (x_1,x_2,b_1,b_2)$ and the Sudakov form factors 
$S_B (t_i)$ and $S_\pi (t_i)$ are given in the appendix 
\ref{ap:FormulasHardPart}.
The operators $O_5$, $O_6$, $O_7$, and $O_8$ have a structure of 
$(V-A) (V+A)$. The sum of their amplitudes is
\begin{eqnarray}
F_e^{P}&=& -32 \pi C_F m_B^2 r_\pi
\int_{0}^{1}d x_{1}d x_{2}\,\int_{0}^{\infty} b_1d b_1 b_2d b_2\,
\phi_B(x_1,b_1)
\nonumber \\
& &\times \left \{ \left[\phi_\pi(x_2,b_2) 
+ (2+x_2) \phi'_\pi(x_2,b_2) r_\pi\right] \alpha_{s}(t_e^1)
h_e (x_1,x_2,b_1,b_2) \right.
\nonumber \\
& &\times 
\exp[-S_{B}(t_e^1)-S_\pi^1(t_e^1)]+ \left[ x_1 \phi_\pi(x_2,b_2) 
+ 2(1-x_1) \phi'_\pi(x_2,b_2) r_\pi\right]
\nonumber \\
&&\times \left . \alpha_{s}(t_e^2)
h_e (x_2,x_1,b_2,b_1)\exp[-S_{B}(t_e^2)-S_\pi^1(t_e^2)] \right \}\;.
\label{bp}
\end{eqnarray}
They are proportional to the factor $r_\pi$.
There are also factorizable annihilation diagrams (g) and (h), where the
$B$ meson can be factored out. 
For the $(V-A)(V-A)$ operators, their contributions always cancel 
between diagram (g) and (h). 
But for the $(V-A)(V+A)$ operators, their contributions are sum
of diagram (g) and (h). 
\begin{eqnarray}
F_a^P&=& - 64 \pi C_F m_B^2 r_\pi
\int_{0}^{1}d x_{2}d x_{3}\,\int_{0}^{\infty} b_2d b_2 b_3d b_3
  \alpha_s (t_{a})h_{a}(x_2,x_3,b_2,b_3)
\nonumber \\
& &\times \left[2 \phi_\pi(x_2,b_2)\phi'_\pi(x_3,b_3)
+x_2 \phi_\pi(x_3,b_3)\phi'_\pi(x_2,b_2) \right] 
  \exp[-S_{\pi}^1(t_a)-S_\pi^2(t_a)] \;,
\label{g}
\end{eqnarray}
These two diagrams can be cut in the middle of the diagrams. 
They provide the main strong phase for non-leptonic $B$ decays.
Note that $F_a^P$ vanishes in the limit of $m_0 = 0$.
So the $m_0$ term in the pion wave function does not only have much
effect on the branching ratios, but also the CP asymmetries.
Besides the factorizable diagrams, we can also calculate the non-factorizable
diagrams  (c) and (d) and also the
 non-factorizable annihilation diagrams (e) and (f). 
In this case, the amplitudes involve all the three meson wave functions.
The integration over $b_3$ can be performed easily using $\delta$ 
function $\delta (b_3-b_1)$ in diagram (c,d) and  $\delta (b_3-b_2)$
for diagram (e,f).
\begin{eqnarray}
{\cal M}_e &=& \frac{32}{3}\pi C_F \sqrt{2N_c} m_B^2
\int_{0}^{1}d x_{1}d x_{2}\,d x_{3}\,\int_{0}^{\infty} b_1d b_1 b_2d b_2\,
\phi_B(x_1,b_1)\phi_\pi(x_2,b_2)
\nonumber \\
& &\times \phi_\pi(x_3,b_1)x_2
\alpha_{s}(t_{d})
 h_{d}(x_1,x_2,x_3,b_1,b_2)\exp[-S_B(t_d)-S_{\pi}^1(t_d)-S_\pi^2(t_d)]\;,
\label{cd}
\\
{\cal M}_a&=& \frac{32}{3}\pi C_F \sqrt{2N_c} m_B^2
\int_{0}^{1}d x_{1}d x_{2}\,d x_{3}\,\int_{0}^{\infty} b_1d b_1 b_2d b_2\,
\phi_B(x_1,b_1) 
\nonumber \\
& &\times  \left\{ -\left[x_2 \phi_\pi(x_2,b_2)\phi_\pi(x_3,b_2)
+(x_2+x_3) \phi'_\pi(x_2,b_2)\phi'_\pi(x_3,b_2) r_\pi^2\right]
 \right.\nonumber \\
&&~~~\times \alpha_{s}(t_{f}^1) h_{f}^{(1)}(x_1,x_2,x_3,b_1,b_2) 
\exp\left[-S_B(t_f^1)-S_{\pi}^1(t_f^1)-S_\pi^2(t_f^1)\right]
\nonumber \\
&& +\left[x_2 \phi_\pi(x_2,b_2)\phi_\pi(x_3,b_2)
+(2+x_2+x_3) \phi'_\pi(x_2,b_2)  \phi'_\pi(x_3,b_2)r_\pi^2\right]
\nonumber \\
&& \times\left.  \alpha_{s}(t_{f}^2)
h_{f}^{(2)}(x_1,x_2,x_3,b_1,b_2)]
\exp[-S_B(t_f^2)-S_{\pi}^1(t_f^2)-S_\pi^2(t_f^2)]
\right \}\;.
\label{e}
\end{eqnarray}
Note that when doing the above integrations over $x_i$ and $b_i$,
we have to include the corresponding Wilson coefficients $C_i$ evaluated
at the corresponding scale $t_i$. 
The expression of Wilson coefficients are channel dependent which are 
shown later in this section.
 The functions $h_i$, coming from the
Fourier transform of 
$H$, are given in Appendix 
\ref{ap:FormulasHardPart}.
In the above equations, we have used the assumption that 
$x_1 \ll x_2,x_3$.
Since the light quark momentum fraction $x_1$ in $B$ meson is peaked at
the small region, while quark momentum fraction $x_2$
 of  pion is peaked 
at $0.5$, this is not a bad approximation. 
After using this approximation, all the diagrams are functions of 
$k_1^-= x_1 m_B/\sqrt{2}$ of $B$ meson only, independent of the variable
of $k_1^+$.
For example, by calculating the diagrams (b) we shall demonstrate it.
\begin{eqnarray}
& & \langle \pi(P_2) \pi(P_3)| O_2^{u\dagger} | B(p_B) \rangle \nonumber \\
&\propto & \int\! d^4k_1 d^4k_2\ \phi_B(k_1)\phi'_\pi(k_2)
\ \frac{q\cdot P_3}{q^2\, \ell^2} \nonumber \\
&=&  \int\! d^4k_1 d^4k_2\ \phi_B(k_1)\phi'_\pi(k_2)
\frac{(P_2^+ - k_1^+)p_B^-}{
\{ 2(P_2^+ - k_1^+)k_1^- + {\bf k}_{1T}^2 \}
\{ 2(k_2^+ - k_1^+)k_1^- + {\bf \ell}_T^2 \}
}  \nonumber \\
&\simeq & \int\! d^4k_1 d^4k_2\ \phi_B(k_1)\phi'_\pi(k_2)
\left\{ 
\frac{p_B^-P_2^+}{(2P_2^+k_1^- + {\bf k}_{1T}^2)
(2k_1^-k_2^+ + {\bf \ell}_T^2)}
+ {\cal O}\left(\frac{\Lambda_{{\rm QCD}}}{m_B^2}\right)
\right\},
\end{eqnarray}
where the momenta are assigned in Figure \ref{fig:diagrams}.
The calculation from second formula to last one is approximated as 
$\langle k_1 \rangle \ll \langle k_2\rangle$. 
This approximation is equal to take the momenta of 
spectator quark in the $B$ meson as $k_1= (0,k_1^-,{\bf k}_{1T})$.
We neglect the last term which is higher order one in terms of $1/m_B$
expansion.
Therefore the integration of eq.(\ref{int}) is performed safely.
Though we calculated the above factorization formulae by one order 
in terms of $\alpha_s$, the radiative corrections at the next order
would emerge in forms of $\alpha_s^2 \ln (m/t)$, where $m$'s denote some
scales, i.e. $m_B$, $1/b, \dots$, in the hard part $H(t)$.
Selecting $t$ as the largest scale in $m$'s, the largest logarithm in
the next order corrections is killed.
Accordingly,
the scale $t_i$'s in the above equations are chosen as
\begin{eqnarray}
t_{e}^1 &=& {\rm max}(\sqrt{x_2} m_B,1/b_1,1/b_2)\;,\\
t_{e}^2 &=& {\rm max}(\sqrt{x_1}m_B,1/b_1,1/b_2)\;,\\
t_{d} &=& {\rm max}(\sqrt{x_1x_2}m_B,
\sqrt{x_2x_3} m_B,1/b_1,1/b_2)\;,\nonumber\\
t_{f}^1 &=& {\rm max}(\sqrt{x_2x_3}m_B,
1/b_1,1/b_2)\;,\nonumber\\
t_{f}^2 &=& {\rm max}(\sqrt{x_2x_3}m_B,\sqrt{x_2+x_3-x_2 x_3} m_B,
1/b_1,1/b_2)\;,\nonumber\\
t_{a} &=& {\rm max}(\sqrt{x_2}m_B, 1/b_2,1/b_3)\;.
\end{eqnarray}
They are given the maximum values of the scales appeared in each diagram.

In the language of the above matrix elements for different 
diagrams eq.(\ref{a}-\ref{e}),
the decay amplitude for $B^0\to \pi^+\pi^-$ can be written as
\begin{eqnarray}
{\cal M} (B^0 \to \pi^+\pi^-) &=& 
f_\pi F_e \left[ \xi_u \left(\frac{1}{3} C_1
+ C_2\right)-\xi_t 
\left(C_4+\frac{1}{3}C_3 +C_{10}+\frac{1}{3}C_9\right)\right]\nonumber\\
&-& f_\pi F_e^P \xi_t \left[ 
C_6+\frac{1}{3}C_5 +C_{8}+\frac{1}{3}C_7\right]\nonumber\\
&+& {\cal M}_e \left[\xi_u C_1-\xi_t (C_3+C_9)\right]\nonumber\\
&+& {\cal M}_a  \left[\xi_u C_2- \xi_t \left( C_3+2C_4 +2C_6+\frac{1}{2}C_8
-\frac{1}{2}C_9+ \frac{1}{2}C_{10} \right)\right]
\nonumber\\
&-& f_B F_a \xi_t\left[  \frac{1}{3} C_5+ C_6
-\frac{1}{6} C_7 - \frac{1}{2} C_8 \right],\label{pm}
\end{eqnarray}
where $\xi_u = V_{ub}^*V_{ud}$, $\xi_t = V_{tb}^*V_{td}$.
The decay width is expressed as 
\begin{equation}
 \Gamma = \frac{G_F^2 m_B^3}{128\pi} |{\cal M}|^2 .
\end{equation}
The $C_i's$ should be calculated at the appropriate scale $t_i$ using 
eq.(\ref{w1},\ref{w2}) in the appendices. 
The decay amplitude of the charge conjugate decay channel 
$\overline B^0\to \pi^+\pi^-$ is the same as eq.(\ref{pm}) except replacing
the CKM matrix elements $\xi_u$ to $\xi_u^*$ and $\xi_t$ to $\xi_t^*$
under the phase convention $CP|B^0\rangle = |\bar{B}^0 \rangle$. 

The decay amplitude for $B^0\to \pi^0\pi^0$ can be written as
\begin{eqnarray}
-\sqrt{2}{\cal M} (B^0 \to \pi^0\pi^0) &=& f_\pi F_e \left[ 
\xi_u \left(C_1+ \frac{1}{3}C_2\right)\right. \nonumber\\
&&\left. +\xi_t \left
(\frac{1}{3}C_3 +C_4 +\frac{3}{2}C_7+\frac{1}{2}C_8-\frac{5}{3}C_9
-C_{10}\right)\right]\nonumber\\
&+& f_\pi F_e^P \xi_t \left[ 
C_6+\frac{1}{3}C_5 -\frac{1}{6}C_7-\frac{1}{2}C_{8}\right]\nonumber\\
&+& {\cal M}_e \left[\xi_u C_2-\xi_t (-C_3+\frac{3}{2}C_8+\frac{1}{2}C_9
+\frac{3}{2}C_{10}
)\right]\nonumber\\
&-& {\cal M}_a \left[\xi_u C_2- \xi_t\left( C_3+2C_4 +2C_6+\frac{1}{2}C_8
-\frac{1}{2}C_9+ \frac{1}{2}C_{10}\right) \right]
\nonumber\\
&+& f_B F_a \xi_t\left[  \frac{1}{3} C_5+ C_6
-\frac{1}{6} C_7 - \frac{1}{2} C_8 \right].\label{00}
\end{eqnarray}

The decay amplitude for $B^+\to \pi^+\pi^0$ can be written as
\begin{eqnarray}
\sqrt{2}{\cal M} (B^+ \to \pi^+\pi^0) &=& f_\pi F_e \left[ \frac{4}{3}
\xi_u (C_1+ C_2)-\xi_t 
\left(2C_{10}+2C_9-\frac{3}{2}C_7-\frac{1}{2}C_8\right)\right]\nonumber\\
&-& f_\pi F_e^P \xi_t \left[ 
\frac{3}{2}C_{8}+\frac{1}{2}C_7\right]\nonumber\\
&+& {\cal M}_e \left[\xi_u (C_1+C_2)-\frac{3}{2}
\xi_t (C_8+C_9+C_{10})\right].\label{p0}
\end{eqnarray}
 From the above equations (\ref{pm},\ref{00},\ref{p0}), it is easy to see that
we have the exact Isospin relation for the three decays:
\begin{equation}
{\cal M} (B^0 \to \pi^+\pi^-) 
-\sqrt{2}{\cal M} (B^0 \to \pi^0\pi^0) =\sqrt{2}{\cal M} (B^+ \to \pi^+\pi^0).
\end{equation}

\section{Numerical calculations and discussions of Results}
\label{sc:NumCalc}

In the numerical calculations we use \cite{pdg}
\begin{eqnarray}
 \Lambda_{\overline{{\rm MS}}}^{(f=4)} = 0.25 \mbox{ GeV,} ~~f_\pi = 0.13
 \mbox{ GeV,} ~~f_B = 0.19  \mbox{ GeV,} \nonumber \\
M_B = 5.2792 \mbox{ GeV,} ~~M_W = 80.41\mbox{ GeV,} \nonumber  \\
\tau_{B^\pm}=1.65\times 10^{-12}\mbox{ s,} ~~ 
  \tau_{B^0}=1.56\times 10^{-12}\mbox{ s}
\label{eq:parm1}
\end{eqnarray}
and
\begin{equation}
m_u=4.5 \mbox{ MeV,} ~~m_d=1.8 m_u,
\label{eq:parm2}
\end{equation}
which is relevant to taking $m_0=1.5$ GeV.
For the $\pi$ wave function, we neglect the $b$ dependence part, which is not
important in numerical analysis. 
We use 
\begin{eqnarray}
\phi_\pi(x) &=& \frac{3}{\sqrt{2N_c}}
 f_\pi  x (1-x)[1+a^A(5(1-2x)^2-1)]  ,
\label{phipa}
\end{eqnarray}
with $a^A=0.8$, which is close to the Chernyak-Zhitnitsky (CZ) wave function
\cite{Chernyak:1984ee}.
For this axial vector wave function 
the asymptotic wave function \cite{asm}, $a^A \sim 0$ , 
is suggested from QCD sum rules~\cite{Bakulev:1996ck},
 diffractive dissociation of high momentum pions~\cite{Ashery:1999nq},
the instanton model~\cite{Petrov:1999kg}, and pion distribution 
functions~ \cite{Jakob:1996hd}, etc., 
but we adopt $a^A=0.8$ according to the discussion in ref.~\cite{kls2}.
$\phi'_\pi$ is  chosen as asymptotic wave function
\begin{eqnarray}
\phi'_\pi(x) &=& \frac{3}{\sqrt{2N_c}}
 f_\pi  x (1-x)[1+a^p(5(1-2x)^2-1)]  ,
\label{phipp}
\end{eqnarray}
with $a^P=0$.
For $B$ meson, the wave function is chosen as 
\begin{eqnarray}
\phi_B(x,b) &=& N_B x^2(1-x)^2 \exp \left
 [ -\frac{M_B^2\ x^2}{2 \omega_{b1}^2} -\frac{1}{2} (\omega_{b2} b)^2\right],
\label{phib}
\end{eqnarray}
with  $\omega_{b1}=\omega_{b2}=0.4$ GeV \cite{bsw}, and 
$N_B=91.745$ GeV is the normalization constant.
In this work, we set $\omega_{b1} = \omega_{b2}$ for simplicity.
We would like to point out that the choice of the meson wave 
functions as in
eqs. (\ref{phipa}-\ref{phib}) and the above parameters can not only explain the experimental 
data of $B\to \pi\pi$, but also $B\to K\pi$ \cite{kls,kls2}, $D\pi$ etc., 
which is the result
of a global fitting. However, since the predicted branching ratio of 
$B\to\pi\pi$ is sensitive to the input parameters $f_B$, $m_0$, $a^A$, $a^P$
and $\omega_{b1}$, we will at first give the numerical results with the above parameters, then we give the allowed parameter regions
of $f_B$, $m_0$, $a^A$, $a^P$ and $\omega_{b1}$ constrained
by the experimental data of $B\to\pi^+\pi^-$ presented by CLEO.

The diagrams (a) and (b) in Fig.\ref{fig:diagrams}, calculated in 
eq.(\ref{a}) correspond to the $B\to\pi$
transition form factor $F^{B\pi}(q^2=0)$, where $q=p_B-P_2$.
Our result is $F^{B\pi}(0)=0.25$ to be consistent with QCD sum rule 
one. This implies that PQCD can explain the transition form factor in
the $B$ meson decays, which is different with the conclusion in
ref.\cite{Feldmann:1999sm}. In that paper, because $m_0$ was not considered, 
perturbative contributions to $F^{B\pi}(0)$ were predicted to be 
much smaller than nonperturbative ones.

Although we take the CZ like wave function ($a^A = 0.8$) for $\phi_\pi$, 
one finds that the above parameters give the pion electromagnetic form
factor to be consistent with the experimental data.
The pion electromagnetic form factor $F_{\pi}(Q^2)$ in PQCD is given as 
\cite{Geshkenbein:1982zs,AISKU}
\begin{eqnarray}
F_\pi (Q^2) &=& 16 \pi C_F \int_0^1\!\!\! dx_2 dx_3
 \int_0^\infty\!\!\!\!\!  b_2 db_2\, b_3 db_3\ 
 \alpha_s(t)\ h_e(x_3,x_2,b_3,b_2) \nonumber \\
& & \times \Bigl\{
 x_2 Q^2  \phi_\pi(x_2,b_2) \phi_\pi(x_3,b_3)
 + 2 m_0^2 (1-x_2)
 \phi_\pi'(x_2,b_2) \phi_\pi'(x_3,b_3) \Bigr\} \nonumber \\
& & \times \exp[ -S_\pi^1(t) -S_\pi^2(t) ],
 \label{eq:piform1}
\end{eqnarray}
where $-Q^2$ is the momentum transfer in this system, 
the scale $t$ is chosen as $t = {\rm max}(\sqrt{x_2} Q,1/b_2,1/b_3)$, 
and $m_B$'s are replaced by $Q$ in the $h_e, S_\pi^1$ and $S_\pi^2$.
One may suspect that around $x_1, x_2 \sim 0$, the gluon and virtual quark
propagators give rise to IR divergences which can not be canceled by
the wave functions. However, in PQCD, the transverse momenta $k_T$
save perturbative calculations from the singularities 
around $x_{1,2} \sim 0$.
There are still IR divergences around $k_T \sim 0$, 
but the Sudakov factor which can be calculated from QCD corrections 
does suppress such a region, i.e., 
non-perturbative contributions, sufficiently.
We show the $Q^2$ dependence of $F_\pi(Q^2)$ (eq.(\ref{eq:piform1})) 
in Figure \ref{fig:Q2dep} with the experimental data \cite{Bebek}.
This figure shows that the parameters we used don't conflict with 
the data.
We also show $F_\pi(Q^2)$ for $a^A = 0.8, 0.4,$ and $0$.
It indicates that $F_\pi(Q^2)$ is fairly insensitive to $a^A$.

The CKM parameters we used here are 
\begin{equation}\begin{array}{ll}
|V_{ud}|=0.9740\pm 0.0010, &
|V_{ub}/V_{cb}|=0.08\pm 0.02, \\
|V_{cb}|=0.0395\pm 0.0017,&
|V_{tb}^*V_{td}|=0.0084\pm 0.0018.
\end{array}
\end{equation}
We leave the CKM angle $\phi_2$ as a free parameter. 
$\phi_2$'s definition is \cite{san}
\begin{equation}
\phi_2=arg\left[-\frac{V_{td}V_{tb}^*}{V_{ud}V_{ub}^*}\right].
\end{equation}
In this parameterization, the decay amplitude of $B\to \pi\pi$ can be
written as 
\begin{eqnarray}
{\cal M} &=& V_{ub}^*V_{ud} T -V_{tb}^* V_{td} P\nonumber\\
 &=& V_{ub}^*V_{ud} T 
\left[1 +z e^{i(\phi_2+\delta)} \right],\label{m}
\end{eqnarray}
where $z=\left|\frac{V_{tb}^* V_{td}}{ V_{ub}^*V_{ud} } \right| 
\left|\frac{P}{T} \right|$, and 
$\delta$ is the relative strong phase between tree (T) diagrams and 
penguin diagrams (P). $z$ and $\delta$ can be calculated from PQCD.
For example, in $B^0 \to \pi^+\pi^-$ decay, we get $z=30$\%, and 
$\delta=130^\circ$, if we use the above parameters.
Here in PQCD
approach, the strong phases come from the non-factorizable diagrams and 
annihilation type diagrams (see (c) $\sim$ (h) 
in Figure \ref{fig:diagrams}). 
The internal quarks and gluons can be on mass shell providing the 
strong phases.
This can also be seen from eq.(\ref{hd}-\ref{hgh}), where the
the modified Bessel function $K_0 (-i f)$ has imaginary part.
Numerical analysis also shows that
 the main contribution to the relative strong phase $\delta$ 
comes from the annihilation diagrams, (g) and (h) 
in Figure \ref{fig:diagrams}.
 From the figure, we can see that they are factorizable diagrams.
$B$ meson annihilates to $ q \bar q$ quark pair then decays to $\pi \pi$
final states. 
The intermediate $ q \bar q$ quark pair represent a number of resonance 
states, which implies final state interaction.
In perturbative calculations, 
the two quark lines can be cut providing the imaginary part. 
The importance of these diagrams also makes the contribution of 
penguin diagrams
more important than previously expected.

This
mechanism of producing CP violation strong phase 
is very different from the so-called
Bander-Silverman-Soni (BSS) mechanism \cite{bss}, where
the strong phase comes from the perturbative penguin diagrams. 
The contribution of BSS
mechanism to the direct CP violation in $B\to \pi^+\pi^-$ is only in the order
of few percent \cite{akl2,neubert}. 
It is higher order corrections ($\alpha_s$ suppressed) in our PQCD approach.
Therefore in our approach we can safely neglect
this contribution.
The corresponding charge conjugate $\bar B$ decay  is 
\begin{eqnarray}
\overline{\cal M} &=& V_{ub}V_{ud}^* T -V_{tb} V_{td}^* P\nonumber\\
 &=& V_{ub}V_{ud}^* T 
\left[1 +z e^{i(-\phi_2+\delta)} \right].\label{mb}
\end{eqnarray}
Therefore the averaged branching ratio for $B\to \pi\pi$ is
\begin{eqnarray}
Br&=& (|{\cal M}|^2 +|\overline{\cal M}|^2)/2\nonumber\\
&=&  \left| V_{ub}V_{ud}^* T \right| ^2
\left[1 +2 z\cos \phi_2 \cos \delta +z^2 \right].\label{br}
\end{eqnarray}
 From this equation, we know that the averaged branching ratio is a 
function of CKM angle $\phi_2$, if $z\cos\delta \neq 0$. 

The averaged branching ratio of $B^0\to \pi^+\pi^-$ decay 
which is predicted from the formulae in the previous section 
is shown as a function of $\phi_2$ in Figure~\ref{bran}.
To consider $m_0$ required from chiral symmetry is essentially different 
with previous paper \cite{kroll}.
This figure shows that $m_0$ enhances the branching ratio to agree with
the experimental data.
There is a significant dependence on the CKM angle $\phi_2$.
The branching ratio of $B^0\to \pi^+ \pi^-$ is larger when $\phi_2$ is larger.
The reason is that the penguin contribution is not small.
The CLEO measured  branching ratio of $B\to \pi^+\pi^-$ \cite{cleo}
\begin{equation}
 {\rm Br}(B \to \pi^+ \pi^-) = (4.3 ^{+1.6}_{-1.4} \pm 0.5)
 \times 10^{-6},\label{cleodata}
\end{equation}
is in good agreement with our predictions.
This prefer a lower value of $\phi_2$.
However, the predicted branching ratio is sensitive to the parameters of input.
Especially it is sensitive to $f_B$, $m_0$ and the meson wave functions. 
Therefore, it is unlikely to use this single channel to determine the 
CKM angle $\phi_2$.

The branching ratios of $B\to\pi\pi$ 
are sensitive to some input parameters.
We give the parameter regions allowed by the experimental data in
eq.~(\ref{cleodata}).
Rerevant parameter are $m_0$, $a^P$, and $\omega_{b1}$. 
Others are specified in the begining of this section.
Here we check the sensitivity of our calculation on parameter
$m_0$, $a^P$, and $\omega_{b1}$.
First we fix $m_0 = 1.5$ GeV and 
show the allowed region for $a^P$ and $\omega_{b1}$. 
This is shown in Figure~\ref{fig:region_pipi}(a).
One finds that the branching ratio is fairly insensitive to $a^P$.
Second we fix $a^P = 0$ and show the allowed region for $\omega_{b1}$ 
and $m_0$. 
This is shown in Figure~\ref{fig:region_pipi}(b).
We see that the allowed region for $\omega_{b1}$ and $m_0$ 
is quite large.
The dependence on $a^A$ for the branching ratio of $B \to \pi^+\pi^-$ 
is given in Figure~\ref{fig:varyCa}. 
As discussed in ref.~\cite{kls2}, the central value of the experimental
data 
$R_D = {\rm Br}(B^- \to D^0\pi^-)/{\rm Br}(\overline{B}^0_d \to
D^+\pi^-)$ requires 
$a^A = 0.8$, but this figure indicates that $B \to \pi^+\pi^-$ decay mode
gives no significant restriction on $a^A$.
Therefore, these figures show that the above set of parameters we choose
for Figure~\ref{bran} is in the
allowed region, and that parameter space producing the experimental 
data, eq.~(\ref{cleodata}), is quite large.

The branching ratio of  $B^+\to \pi^+ \pi^0$ has little dependence 
on $\phi_2$. 
It is easy to understand  since there is only one dominant contribution
 from tree diagrams. The QCD penguin contribution is canceled by isospin 
relation
and the electroweak contribution is very small giving only a slight 
dependence on $\phi_2$. 
The branching ratio of this decay is predicted as $ 3 \times 10^{-6}$,
using the parameters we list in the beginning of this section.

For the decay of  $B^0\to \pi^0 \pi^0$, the situation is similar to 
that of $B^0\to \pi^+\pi^-$.  There
are large contributions from both tree and penguin diagrams. 
We show the averaged branching ratio of $B^0\to \pi^0 \pi^0$ as a function of 
$\phi_2$ in Figure~\ref{p0p0}.
Although the branching ratio is small, the dependence of $\phi_2$ is
significant.
The predicted branching ratio of $B^0\to \pi^0 \pi^0$ is less than 
$10^{-6}$. 
This is difficult for the B factories to measure the separate 
branching ratios of $B^0$ and $\bar B^0$. 
In this case, the proposed isospin method to measure the CKM angle $\phi_2$
\cite{gro} does not work in the B factories, since it requires the 
measurement of $B^0\to \pi^0 \pi^0$ and $\bar B^0\to \pi^0 \pi^0$.

Using eq.(\ref{m},\ref{mb}),
the direct CP violating parameter is
\begin{eqnarray}
A_{CP}^{dir} &=& \frac{|{\cal M}|^2 -|\overline{\cal M}|^2}{
|{\cal M}|^2 +|\overline{\cal M}|^2}\nonumber\\ \label{dir}
&=& \frac{-2 z \sin \phi_2 \sin\delta}{1 + 2 z\cos \phi_2 \cos \delta +z^2}.
\end{eqnarray}
The direct CP asymmetry is nearly proportional to $\sin \phi_2$. 
We show
the direct CP violation parameters (percentage) as a function of $\phi_2$ 
in figure~\ref{cppm}.
Unlike the averaged branching ratios, the predicted CP violation in 
B decays does not depend much on the wave functions.
They cancel each between the charge conjugate states 
shown in the above equation.
The direct CP violation parameter of $B^0\to \pi^+ \pi^-$ and 
$\pi^0 \pi^0$ can be very large, which can be as large as 40\%, and 20\%
 when $\phi_2$ is near $70^\circ$. 
Because there is no annihilation diagram contribution in $B^+\to \pi^+ \pi^0$, 
the penguin contribution is negligible. 
The direct CP violation parameter of $B^+\to \pi^+ \pi^0$ is also very small.
 It is a horizontal line in Figure~\ref{cppm}.

For the neutral $B^0$ decays, there is more complication from the 
$B^0$-$\overline {B^0}$ mixing. 
The CP asymmetry is time dependent \cite{akl2,wu}:
\begin{equation}
A_{CP} (t) \simeq A_{CP}^{dir} \cos (\Delta mt) + a_{\epsilon + \epsilon '}
\sin (\Delta m t),
\end{equation}
where $\Delta m$ is the mass difference of the two  mass eigenstates of
neutral $B$ mesons. 
The direct CP violation parameter $A_{CP}^{dir}$ is
already defined in eq.(\ref{dir}). While the mixing-related CP violation 
parameter is defined as
\begin{equation}
a_{\epsilon +\epsilon'}=\frac{ -2Im (\lambda_{CP})}
{1+|\lambda_{CP}|^2},
\end{equation}
where 
\begin{equation}
\lambda_{CP} = \frac{ V_{tb}^*V_{td} \langle f |H_{eff}| \overline B^0\rangle}
{ V_{tb}V_{td}^* \langle f |H_{eff}| B^0\rangle}.
\end{equation}
Using equations (\ref{m},\ref{mb}), we can derive as
\begin{equation}
\lambda_{CP} = e^{2i\phi_2}\frac{ 1+ze^{i(\delta-\phi_2)} }{ 
 1+ze^{i(\delta+\phi_2)} }.
\end{equation}
Usually, people believe that the penguin diagram contribution is suppressed
comparing with the tree contribution, i.e. $z \ll 1$. Such that 
$\lambda_{CP} \simeq \exp [2i \phi_2]$, $a_{\epsilon+\epsilon'} = 
-\sin 2 \phi_2$, and $A_{CP}^{dir}\simeq 0$. 
That is the previous idea of extracting $\sin 2\phi_2 $ 
 from the CP measurement of $B^0 \to \pi^+ \pi^-$. 
However, $z$ is not very small.
 From Figure~\ref{ae}, we can see that $a_{\epsilon+\epsilon'} $
is not a simple  $-\sin 2 \phi_2$ behavior due to the so called 
penguin pollution.

If we integrate the time variable $t$, we will get the total CP asymmetry as
\begin{equation}
A_{CP} = \frac{1}{1+x^2} A_{CP}^{dir} + \frac{x}{1+x^2}a_{\epsilon +\epsilon'},
\end{equation}
with $x=\Delta m/\Gamma \simeq 0.723 $ for the $B^0$-$\overline B^0$ mixing
in SM \cite{pdg}. 

The integrated CP asymmetries of $B^0\to \pi^+ \pi^-$ and
$B^0\to \pi^0 \pi^0$ are shown in Figure~\ref{acp}.
 Unlike the averaged branching ratios, the CP asymmetry is not sensitive 
to the wave functions, since these parameter dependences canceled out.
 It is rather stable.
If we can measure the integrated CP asymmetry from the experiments, then we
can use this figure to determine the value of $\phi_2$.

\section{Summary}

We performed the calculations of $B^0\to \pi^+\pi^-$, $B^+\to \pi^+\pi^0$, 
and $B^0\to \pi^0\pi^0$ decays, in a perturbative QCD approach.
In this approach, we  calculate the non-factorizable contributions and 
annihilation type contributions in addition to the usual factorizable
contributions. The predicted branching ratios of $B^0\to \pi^+\pi^-$
is in good agreement with the experimental measurement by the
 CLEO Collaboration.

We found that the annihilation contributions were not as 
small as expected in a simple argument. 
The annihilation diagram, which provides the dominant strong phases,
plays an important role in the CP violation asymmetries.
We expect large direct CP asymmetries in the decay of $B^0\to \pi^+\pi^-$,
and $B^0\to \pi^0\pi^0$. 
The ordinary method of measuring the
 CKM angle $\phi_2$ will suffer from the large 
penguin pollution. 
The isospin method does not help, since the B factories can not
measure well the small branching ratio of $B^0 \to \pi^0 \pi^0$.
Working in our PQCD approach, we give the predicted dependence of 
CP asymmetry on CKM angle $\phi_2$. 
Using this dependence, the current running B factories in KEK and SLAC
will be able to measure the CKM angle $\phi_2$.

\section*{Acknowledgments}

We thank our PQCD group members: Y.Y. Keum, E. Kou, T. Kurimoto, H.N. Li,
T. Morozumi, 
A.I. Sanda, N. Sinha, R. Sinha, and T. Yoshikawa
 for helpful discussions.
This work was supported by the Grant-in-Aid for Scientific Research 
on Priority Areas (Physics of CP violation).
C.D. L. and M.Z. Y. thanks JSPS for support.
K. U. thanks JSPS for partial support.

\begin{appendix}

\section{Wilson Coefficients}
In this appendix we present the weak effective Hamiltonian
${\cal H}_{{\it eff}}$ which we used to calculate the hard part $H(t)$
in eq.(\ref{eq:factorization_formula}).
The ${\cal H}_{{\it eff}}$ for the $\Delta B=1$ transitions at the 
scale smaller than $m_W$ is given as
\begin{equation}
\label{heff}
{\cal H}_{{\it eff}}
= \frac{G_{F}} {\sqrt{2}} \, \left[ V_{ub} V_{ud}^*
\left (C_1 O_1^u + C_2 O_2^u \right)
-
V_{tb} V_{td}^* \,
\left(\sum_{i=3}^{10}
C_{i} \, O_i + C_g O_g \right) \right] \quad .
\end{equation}
We
specify below the operators in  ${\cal H}_{{\it eff}}$ for $b \to d$:
\begin{equation}\begin{array}{llllll}
 O_1^{u} & = &  \bar d_\alpha\gamma^\mu L u_\beta\cdot \bar 
u_\beta\gamma_\mu L b_\alpha\ ,
&O_2^{u} & = &\bar d_\alpha\gamma^\mu L u_\alpha\cdot \bar 
u_\beta\gamma_\mu L b_\beta\ , \\ 
O_3 & = & \bar d_\alpha\gamma^\mu L b_\alpha\cdot \sum_{q'}\bar
 q_\beta'\gamma_\mu L q_\beta'\ ,   &
O_4 & = & \bar d_\alpha\gamma^\mu L b_\beta\cdot \sum_{q'}\bar 
q_\beta'\gamma_\mu L q_\alpha'\ , \\
O_5 & = & \bar d_\alpha\gamma^\mu L b_\alpha\cdot \sum_{q'}\bar 
q_\beta'\gamma_\mu R q_\beta'\ ,   &
O_6 & = & \bar d_\alpha\gamma^\mu L b_\beta\cdot \sum_{q'}\bar 
q_\beta'\gamma_\mu R q_\alpha'\ , \\
O_7 & = & \frac{3}{2}\bar d_\alpha\gamma^\mu L b_\alpha\cdot 
\sum_{q'}e_{q'}\bar q_\beta'\gamma_\mu R q_\beta'\ ,   &
O_8 & = & \frac{3}{2}\bar d_\alpha\gamma^\mu L b_\beta\cdot 
\sum_{q'}e_{q'}\bar q_\beta'\gamma_\mu R q_\alpha'\ , \\
O_9 & = & \frac{3}{2}\bar d_\alpha\gamma^\mu L b_\alpha\cdot 
\sum_{q'}e_{q'}\bar q_\beta'\gamma_\mu L q_\beta'\ ,   &
O_{10} & = & \frac{3}{2}\bar d_\alpha\gamma^\mu L b_\beta\cdot 
\sum_{q'}e_{q'}\bar q_\beta'\gamma_\mu L q_\alpha'~. 
\label{operators}
\end{array}
\end{equation}
Here $\alpha$ and $\beta$ are the $SU(3)$ color indices;
$L$ and $R$ are the left- and right-handed projection operators with
$L=(1 - \gamma_5)$, $R= (1 + \gamma_5)$.
The sum over $q'$ runs over the quark fields that are active at the scale 
$\mu=O(m_b)$, i.e., $(q'\epsilon\{u,d,s,c,b\})$. 

The PQCD approach works well for the leading twist approximation and leading 
double logarithm summation. For the Wilson coefficients, we will also use 
the leading logarithm summation for the QCD corrections, although the 
next-to-leading order calculations already exist in the literature \cite{buras}. 
This is the consistent way to cancel the explicit $\mu$ dependence in the
theoretical formulae. 

At $m_W$ scale, the Wilson coefficients are evaluated for leading order as:
\begin{eqnarray}
	C_{2w}&=&1,\nonumber\\
	C_{iw}&=&0, ~~~~i=1,8,10,\nonumber\\
	C_{3w}&=&-\frac{\alpha_s(m_W)}{24\pi} E_0+\frac{\alpha}{6\pi} 
\frac{1}{\sin^2 \theta_W} (2 B_0+ C_0),\nonumber\\
	C_{4w}&=&\frac{\alpha_s(m_W)}{8\pi} E_0,\nonumber\\
	C_{5w}&=&-\frac{\alpha_s(m_W)}{24\pi} E_0,\nonumber\\
	C_{6w}&=&\frac{\alpha_s(m_W)}{8\pi} E_0,\nonumber\\
	C_{7w}&=&\frac{\alpha}{6\pi} (4 C_0+D_0),\nonumber\\
	C_{9w}&=&\frac{\alpha}{6\pi} \left[4 C_0+D_0+\frac{1}{\sin^2 \theta_W}
  (10 B_0-4 C_0)\right],
\end{eqnarray}
where
\begin{eqnarray}
	B_0&=&\frac{1}{4}\left(\frac{x}{1-x}+\frac{x}{(x-1)^2}\ln x
\right), \nonumber\\
	C_0&=& \frac{x}{8} \left(\frac{x-6}{x-1}+
\frac{3 x+2}{(x-1)^2}\ln x\right),\nonumber\\
D_0 &=& -\frac{4}{9} \ln x+\frac{-19 x^3+25 x^2}{36(x-1)^3}
         +\frac{x^2 (5 x^2-2 x-6)}{18(x-1)^4}\ln x,\nonumber\\
E_0 &=& -\frac{2}{3} \ln x +\frac{x(x^2+11 x-18)}{12(x-1)^3}
         +\frac{x^2 (4 x^2-16 x+15)}{6(x-1)^4}\ln x ,
\end{eqnarray}
with	$x=m_t^2/m_W^2$.

If the scale $ m_b< t< m_W$, then we evaluate the Wilson coefficients at $t$
scale using leading logarithm running equations (\ref{w1}), 
in the appendix \ref{ap:above_mb}.
In numerical calculations, we use
  $\alpha_s=
4\pi/[\beta_1 \ln(t^2/{\Lambda_{QCD}^{(5)}}^2)]$ which is the leading order
expression with $\Lambda_{QCD}^{(5)}=193$MeV, derived
 from $\Lambda_{QCD}^{(4)}=250$MeV. 
Here $\beta_1=(33-2n_f)/3$, with the 
appropriate number of active quarks $n_f$. 
$n_f=5$ when scale $t$ is larger than $m_b$.

The Wilson coefficients evaluated at $t=m_b=4.8$GeV scale using the above
equations are
\begin{equation}
\begin{array}{ll}
	C_1= -0.27034, &
	C_2=  1.11879,\\
	C_3=  0.01261, &
	C_4= -0.02695,\\
	C_5=  0.00847, &
	C_6= -0.03260,\\
	C_7=  0.00109,&
	C_8=  0.00040,\\
	C_9= -0.00895,&
	C_{10}= 0.00216.
\end{array}\label{ci}
\end{equation}

If the scale $t < m_b$, then we evaluate the Wilson coefficients at $t$
scale using the input of eq.(\ref{ci}), and the  formulae in appendix 
\ref{ap:below_mb} for
four active quarks ($n_f=4$) (again in leading logarithm approximation).

\section{Formulas for the hard part calculations}
\label{ap:FormulasHardPart}

In this appendix we present the explicit expression of the formulas 
we used in section 3. First, we show 
the exponent $s(k,b)$
appearing in eq.~(\ref{wp}-\ref{Se}).  It is given, in terms of
the variables,
\begin{eqnarray}
{\hat q} \equiv  {\rm ln}\left(k/\Lambda\right),\;\;\;\;\;
{\hat b} \equiv {\rm ln}(1/b\Lambda)
\end{eqnarray}
by
\begin{eqnarray}
s(k,b)&=&\frac{2}{3\beta_{1}}\left[\hat{q}\ln\left(\frac{\hat{q}}
{\hat{b}}\right)-
\hat{q}+\hat{b}\right]+
\frac{A^{(2)}}{4\beta_{1}^{2}}\left(\frac{\hat{q}}{\hat{b}}-1\right)
\nonumber \\
& &-\left[\frac{A^{(2)}}{4\beta_{1}^{2}}-\frac{1}{3\beta_{1}}
\left(2\gamma_E-1-\ln 2\right)\right]
\ln\left(\frac{\hat{q}}{\hat{b}}\right).
\label{sss}
\end{eqnarray}
The above coefficients $\beta_{1}$ and $A^{(2)}$ are
\begin{eqnarray}
& &\beta_{1}=\frac{33-2n_{f}}{12}
\nonumber \\
& & A^{(2)}=\frac{67}{9}-\frac{\pi^{2}}{3}-\frac{10}{27}n_
{f}+\frac{8}{3}\beta_{1}\ln\left(\frac{e^{\gamma_E}}{2}\right)\; ,
\end{eqnarray}
where $\gamma_E$ is the Euler constant. 

Note that $s$ is defined for ${\hat q}\ge {\hat b}$, and set to zero for
${\hat q}<{\hat b}$. As a similar treatment, the complete Sudakov factor
$\exp(-S)$ is set to unity, if $\exp(-S)>1$, in the numerical analysis.
This corresponds to a truncation at large $k_T$, which spoils the
on-shell requirement for the light valence quarks. The quark lines with
large $k_T$ should be absorbed into the hard scattering amplitude, instead
of the wave functions. 

$e^{-S_B(t)}$, $e^{-S_\pi^1(t)}$, and $e^{-S_\pi^2(t)}$ used in the 
amplitudes are expressions abbreviated to combine the Sudakov factor and 
single ultraviolet logarithms associated with the $B$ and $\pi$ meson
wave functions.
The exponents are defined as
\begin{eqnarray}
S_{B}(t) &=& s\left(x_1 m_B/\sqrt{2}, b_1\right) 
-\frac{1}{\beta_1} \ln\frac{\ln(t/\Lambda)}{-\ln(b_1\Lambda)},
\label{wp}\\
S_{\pi}^1(t) &=& 
 s\left(x_2 m_B/\sqrt{2}, b_2\right) 
+s\left((1-x_2) m_B/\sqrt{2}, b_2\right)
-\frac{1}{\beta_1}\ln\frac{\ln(t/\Lambda)}{-\ln(b_2\Lambda)},
\label{Sc}\\
S_{\pi}^2(t) &=&  s\left(x_3 m_B/\sqrt{2}, b_3\right) 
+s\left((1-x_3) m_B/\sqrt{2}, b_3\right) 
-\frac{1}{\beta_1} \ln\frac{\ln(t/\Lambda)}{-\ln(b_3\Lambda)}
.\label{Se}
\end{eqnarray}
The last term of each equations is the integration result of 
the last term in eq.(\ref{eq:factorization_formula}).

The function $h_i$'s, coming from the
Fourier transform of hard part 
$H$, are given as,
\begin{eqnarray}
h_e(x_1,x_2,b_1,b_2)&=& K_{0}\left(\sqrt{x_1x_2} m_B b_1\right)
 \left[\theta(b_1-b_2)K_0\left(\sqrt{x_2} m_B
b_1\right)I_0\left(\sqrt{x_2} m_B b_2\right)\right.
\nonumber \\
& &\;\;\;\;\left.
+\theta(b_2-b_1)K_0\left(\sqrt{x_2}  m_B b_2\right)
I_0\left(\sqrt{x_2}  m_B b_1\right)\right],
\label{ha}\\
h_{d}(x_1,x_2,x_3,b_1,b_2)&=& K_{0}\left(-i\sqrt{x_2 x_3} m_B b_2\right)
\left[\theta(b_1-b_2)K_0\left(\sqrt{x_1 x_2} m_B
b_1\right)I_0\left(\sqrt{x_1 x_2} m_B b_2\right)\right.
\nonumber \\
& &\left.
+\theta(b_2-b_1)K_0\left(\sqrt{x_1 x_2}  m_B b_2\right)
I_0\left(\sqrt{x_1 x_2}  m_B b_1\right)\right],
\label{hd}\\
h_{f}^{(1)}(x_1,x_2,x_3,b_1,b_2)&=& 
 K_{0}\left(-i\sqrt{x_2 x_3} m_B b_1\right)
 \nonumber \\
& &\times \left[\theta(b_1-b_2)K_0\left(-i\sqrt{x_2 x_3} m_B
b_1\right)J_0\left(\sqrt{x_2 x_3} m_B b_2\right)\right.
\nonumber \\
& &\left.
+\theta(b_2-b_1)K_0\left(-i\sqrt{x_2 x_3}  m_B b_2\right)
J_0\left(\sqrt{x_2 x_3}  m_B b_1\right)\right],
\label{hef} \\
h_{f}^{(2)}(x_1,x_2,x_3,b_1,b_2)&=&
K_{0}\left(\sqrt{x_2+x_3-x_2 x_3} m_B b_1\right)
\nonumber \\
& &\times \left[\theta(b_1-b_2)K_0\left(-i\sqrt{x_2 x_3} m_B
b_1\right)J_0\left(\sqrt{x_2 x_3} m_B b_2\right)\right.
\nonumber \\
& &\left.
+\theta(b_2-b_1)K_0\left(-i\sqrt{x_2 x_3}  m_B b_2\right)
J_0\left(\sqrt{x_2 x_3}  m_B b_1\right)\right],
\label{hf} \\
h_{a}(x_2,x_3,b_2,b_3)&=& K_0 \left(-i\sqrt{x_2 x_3} m_B b_3\right)
\left[\theta(b_2-b_3)K_0\left(-i\sqrt{x_2 } m_B
b_2\right)J_0\left(\sqrt{x_2 } m_B b_3\right)\right.
\nonumber \\
& &\left.~~~~
+\theta(b_3-b_2)K_0\left(-i\sqrt{x_2 }  m_B b_3\right)
J_0\left(\sqrt{x_2 }  m_B b_2\right)\right], \label{hgh}
\end{eqnarray}
with $J_0$ the Bessel function and  $K_0$, $I_0$ modified Bessel functions 
$K_0(-ix)= -(\pi/2) Y_0 (x) + i (\pi/2) J_0 (x)$.

\section{Wilson coefficients running equations above $m_b$ scale}
\label{ap:above_mb}

In this appendix, we list formulas for renormalization group running 
 from $m_W$ scale to $t$ scale, where $t> m_b$. 
These formulas are derived from the leading logarithm QCD 
corrections with five active quarks \cite{buras}.
\begin{eqnarray}
C_1 &=& \frac{1}{2} \left(\eta^{-6/23} - \eta^{12/23} \right),\nonumber\\
  C_2 &=& \frac{1}{2}  \left(\eta^{-6/23} + \eta^{12/23}\right), \nonumber\\
 C_3 &=& 0.0510 \eta^{-0.4086} - 0.0714 \eta^{-6/23} + 
    0.0054 \eta^{-0.1456} \nonumber\\
&&-0.1403 \eta^{0.4230} - 
    0.0113 \eta^{0.8994} + 1/6 \eta^{12/23} \nonumber\\
&&+ 
    C_{3w} (0.2868\eta^{-0.4086} +  0.0491\eta^{-0.1456} + 
       0.6579\eta^{0.4230}  +   0.0061\eta^{0.8994})\nonumber\\
&&+
    C_{4w}(0.3287\eta^{-0.4086}  +   0.0424\eta^{-0.1456} - 
       0.3263\eta^{0.4230}  -   0.0448\eta^{0.8994}) \nonumber\\
&&+ C_{5w}(-0.0629\eta^{-0.4086} + 0.1629\eta^{-0.1456} - 
       0.1846\eta^{0.4230} +  0.0846\eta^{0.8994}) \nonumber\\
&&+ C_{6w} (0.0447\eta^{-0.4086} -  0.0063\eta^{-0.1456} - 
       0.2610\eta^{0.4230}+  0.2226\eta^{0.8994})\nonumber\\
&&+ 
    C_{9w}\left (-0.0325 \eta^{-0.4086} + 0.0357 \eta^{-6/23} - 
       0.0016 \eta^{-0.1456}\right.\nonumber\\
&& \left.~~~~+0.2342 \eta^{0.4230} - 
       0.25 \eta^{12/23} +0.0141 \eta^{0.8994}\right) \nonumber\\
&&+ 
    C_{7w}\left (-0.0063 \eta^{-0.4086}  + 0.0163 \eta^{-0.1456}  - 
       0.0185 \eta^{0.4230} +0.0085 \eta^{0.8994} \right),\nonumber
\\
C_4 &=&  0.0984 \eta^{-0.4086} - 0.0714 \eta^{-6/23} + 
    0.0026 \eta^{-0.1456} \nonumber\\
&&+ 0.1214 \eta^{0.4230} - 
    1/6 \eta^{12/23}+0.0156 \eta^{0.8994}\nonumber\\
&& + C_{3w} (0.5539\eta^{-0.4086} + 0.0239\eta^{-0.1456} - 
       0.5693\eta^{0.4230}  - 0.0085\eta^{0.8994})\nonumber\\
&& + C_{4w} (0.6348\eta^{-0.4086}  +  0.0206\eta^{-0.1456} + 
       0.2823\eta^{0.4230}+  0.0623\eta^{0.8994})\nonumber\\
&& + C_{5w} (-0.1215\eta^{-0.4086}  +  0.0793\eta^{-0.1456} + 
       0.1597\eta^{0.4230}  -  0.1175\eta^{0.8994})\nonumber\\
&& + C_{6w} (0.0864\eta^{-0.4086}  - 0.0031\eta^{-0.1456} + 
       0.2259\eta^{0.4230}  - 0.3092\eta^{0.8994})\nonumber\\
&& + 
    C_{9w}\left (-0.0627 \eta^{-0.4086} + 0.0357 \eta^{-6/23} - 
       0.0008 \eta^{-0.1456} \right.\nonumber\\
&&\left.~~~~- 0.2027 \eta^{0.4230} + 
       0.25 \eta^{12/23} -0.0196 \eta^{0.8994}\right)  \nonumber\\
&&+ 
    C_{7w} \left(-0.0122 \eta^{-0.4086} +0.0079 \eta^{-0.1456}  + 
       0.0160 \eta^{0.4230}  - 0.0117 \eta^{0.8994}\right ),\nonumber\\
C_5 &=&  -0.0397 \eta^{-0.4086}  + 0.0304 \eta^{-0.1456} + 
    0.0117 \eta^{0.4230} - 0.0025 \eta^{0.8994} \nonumber\\
&& +C_{3w} (-0.2233\eta^{-0.4086} +  0.2767\eta^{-0.1456} - 
       0.0547\eta^{0.4230}  + 0.0013\eta^{0.8994}) \nonumber\\
&& +C_{4w} (-0.2559\eta^{-0.4086}+ 
       0.2385\eta^{-0.1456} +  0.0271\eta^{0.4230}  - 
       0.0098\eta^{0.8994}) \nonumber\\
&& + C_{5w} (0.0490\eta^{-0.4086}+ 0.9171\eta^{-0.1456} + 
       0.0154\eta^{0.4230} +  0.0185\eta^{0.8994})\nonumber\\
&& + C_{6w} (-0.0348\eta^{-0.4086} - 0.0357\eta^{-0.1456} + 
       0.0217\eta^{0.4230} + 0.0488\eta^{0.8994})\nonumber\\
&& + 
  C_{9w} \left(0.0253 \eta^{-0.4086}  - 0.0089 \eta^{-0.1456}- 
       0.0195 \eta^{0.4230}  +0.0031 \eta^{0.8994}\right) \nonumber\\
&&  + 
    C_{7w} \left(0.0049 \eta^{-0.4086} \right. \nonumber\\
&&+ \left.0.0917 \eta^{-0.1456} - 
       0.1 \eta^{-3/23} + 0.0015 \eta^{0.4230}  + 
       0.0019 \eta^{0.8994}\right ),\nonumber\\
  C_6 &=& 0.0335 \eta^{-0.4086}  - 0.0112 \eta^{-0.1456} + 
    0.0239 \eta^{0.4230} - 0.0462 \eta^{0.8994}\nonumber\\
&& + C_{3w} (0.1885\eta^{-0.4086} - 0.1017\eta^{-0.1456} - 
       0.1120\eta^{0.4230}  + 0.0251\eta^{0.8994})\nonumber\\
&& + C_{4w} (0.2160\eta^{-0.4086}  -  0.0877\eta^{-0.1456} + 
       0.0555\eta^{0.4230} -  0.1839\eta^{0.8994})\nonumber\\
&& + C_{5w} (-0.0414\eta^{-0.4086} - 0.3370\eta^{-0.1456} + 
       0.0314\eta^{0.4230} +  0.3469\eta^{0.8994}) \nonumber\\
&& + C_{6w} (0.0294\eta^{-0.4086}  + 0.0131\eta^{-0.1456} + 
       0.0444\eta^{0.4230}  +   0.9131\eta^{0.8994})\nonumber\\
&& + 
    C_{9w} \left(-0.0213 \eta^{-0.4086} + 0.0033 \eta^{-0.1456} - 
       0.0399 \eta^{0.4230}  +0.0579 \eta^{0.8994}\right)  \nonumber\\
&& + 
    C_{7w} \left (-0.0041 \eta^{-0.4086} - 0.0337 \eta^{-0.1456} + 
       \eta^{-3/23}/30 +0.0031 \eta^{0.4230} \right .\nonumber\\
&&~~~~+ \left.
       0.0347 \eta^{0.8994} -\eta ^{24/23}/30\right ),
\nonumber\\
 C_7 &=& C_{7w} \eta^{-3/23},
\nonumber\\
  C_8 &=& \frac{1}{3} C_{7w} \left (-\eta^{-3/23} +\eta^{24/23}\right ),
 \nonumber\\
  C_9 &=& \frac{1}{2} C_{9w} \left (  \eta^{-6/23}  +  \eta^{12/23} \right ),
\nonumber\\
 C_{10} &=&\frac{1}{2}  C_{9w} \left (\eta^{-6/23} -  \eta^{12/23} \right ) ,
\label{w1}
\end{eqnarray}
where $\eta=\alpha_s(t)/\alpha_s(m_W)$.

\section{Wilson coefficients running equations below $m_b$ scale}
\label{ap:below_mb}

In this appendix, we list formulas for renormalization group running 
 from $m_b$ scale to $t$ scale, where $t< m_b$. 
These formulas are derived from the leading logarithm QCD 
corrections with four active quarks \cite{buras}.
\begin{eqnarray}
	CC_1&=&\frac{1}{2} C_2 \left ( \zeta^{-6/25} -\zeta^{12/25} \right ) 
        +\frac{1}{2}  C_1 \left (  \zeta^{-6/25} +  \zeta^{12/25}\right ) ,
 \nonumber\\
	CC_2&=& \frac{1}{2} C_2 \left ( \zeta^{-6/25} +  \zeta^{12/25}\right ) + 
         \frac{1}{2} C_1 \left ( \zeta^{-6/25}  - \zeta^{12/25}  \right ),
 \nonumber\\
	CC_3&=& C_{4} \left (0.3606 \zeta^{-0.3469} +  0.03166 \zeta^{-0.1317} - 
       0.3626 \zeta^{0.4201}  - 0.0297 \zeta^{0.8451}\right ) \nonumber\\
&&+ 
      C_{10} \left (0.0149 \zeta^{-0.3469}  - 0.0020 \zeta^{-0.1317} - 
       0.4981 \zeta^{0.4201} + 0.5 \zeta^{12/25}- 0.0148 \zeta^{0.8451}\right )
 \nonumber\\
&&  + 
       C_2 \left (0.0651 \zeta^{-0.3469} -  0.0833 \zeta^{-6/25} \right . 
\nonumber\\
&&+ \left.
       0.0046 \zeta^{-0.1317} - 0.2265 \zeta^{0.4201} + 0.25 \zeta^{12/25}
-     0.0099 \zeta^{0.8451}\right )\nonumber\\
&& + 
       C_{3} \left (0.3308 \zeta^{-0.3469}  + 0.0356 \zeta^{-0.1317}
 +       0.6337 \zeta^{0.4201}  -  0.0001 \zeta^{0.8451}\right ) \nonumber\\
&& + 
       C_1 \left (0.0502 \zeta^{-0.3469} - 0.0833 \zeta^{-6/25} + 
        0.0066 \zeta^{-0.1317}+0.2717 \zeta^{0.4201}\right. \nonumber\\
&&\left. -0.25 \zeta^{12/25}+ 
        0.0049 \zeta^{0.8451}\right ) + 
       C_{9} \left (-0.0149 \zeta^{-0.3469}\right . \nonumber\\
&& \left. + 0.0020 \zeta^{-0.1317} + 
         0.4981 \zeta^{0.4201} - 0.5 \zeta^{12/25} + 0.0148 \zeta^{0.8451}\right )
\nonumber\\
&&+ 
       C_5 \left (-0.0598 \zeta^{-0.3469} + 0.1371 \zeta^{-0.1317} - 
        0.1473 \zeta^{0.4201}+   0.0700 \zeta^{0.8451}\right )  \nonumber\\
&& + 
       C_{6} \left (0.0377 \zeta^{-0.3469} -   0.0045 \zeta^{-0.1317} -  
        0.2210 \zeta^{0.4201}  + 0.18775 \zeta^{0.8451}\right )  \nonumber\\
&&+
       C_{7} \left (-0.0150 \zeta^{-0.3469}  + 0.0343 \zeta^{-0.1317}  - 
        0.0368 \zeta^{0.4201}  +  0.0175 \zeta^{0.8451}\right ) \nonumber\\
&&+ 
       C_8 \left (0.009 \zeta^{-0.3469} -  0.0011 \zeta^{-0.1317} - 
        0.0553 \zeta^{0.4201}  + 0.0469 \zeta^{0.8451} \right ),
 \nonumber\\
	CC_4&=&C_{6} \left (0.0640 \zeta^{-0.3469} - 0.0021 \zeta^{-0.1317} + 
        0.2018 \zeta^{0.4201}  - 0.2637 \zeta^{0.8451}
       \right ) \nonumber\\
&&+ C_5 \left (-0.10156 \zeta^{-0.3469}  +  0.06538 \zeta^{-0.1317} + 
        0.1345 \zeta^{0.4201} -  0.09836 \zeta^{0.8451}\right ) \nonumber\\
&&  + 
       C_{9} \left (-0.02528 \zeta^{-0.3469}  + 0.0009 \zeta^{-0.1317} \right . 
\nonumber\\
&&- \left. 
        0.4549 \zeta^{0.4201} + 0.5 \zeta^{12/25} - 0.0207 \zeta^{0.8451}\right ) 
\nonumber\\
&&+ 
       C_1 \left (0.08515 \zeta^{-0.3469} -  0.0833 \zeta^{-6/25} + 
       0.0031 \zeta^{-0.1317}\right . \nonumber\\
&&\left. ~~~~-0.24809 \zeta^{0.4201}+0.25 \zeta^{12/25}- 
       0.00688 \zeta^{0.8451}\right ) \nonumber\\
&&+ 
       C_{3} \left (0.5615 \zeta^{-0.3469}  +  0.01699 \zeta^{-0.1317} - 
         0.5787 \zeta^{0.4201} +  0.0002 \zeta^{0.8451}\right ) \nonumber\\
&& + 
       C_2 \left (0.1104 \zeta^{-0.3469}-0.0833 \zeta^{-6/25}\right . \nonumber\\
&&\left.
           +0.0022 \zeta^{-0.1317}  +0.2068 \zeta^{0.4201} 
- 0.25 \zeta^{12/25}+  0.0139 \zeta^{0.8451}\right ) \nonumber\\
&& + 
        C_{10} \left (0.0253 \zeta^{-0.3469} - 0.0009 \zeta^{-0.1317} \right . 
\nonumber\\
&&+ \left. 
        0.4549 \zeta^{0.4201} - 0.5 \zeta^{12/25} + 0.0207 \zeta^{0.8451}\right ) 
\nonumber\\
&&+
       C_{4} \left (0.6121 \zeta^{-0.3469} +  0.0151 \zeta^{-0.1317} + 
         0.3311 \zeta^{0.4201} +  0.0417 \zeta^{0.8451}\right ) \nonumber\\
&&   + 
      C_8 \left (0.0160 \zeta^{-0.3469}  - 0.0005 \zeta^{-0.1317}+ 
         0.0505 \zeta^{0.4201}  -  0.0659 \zeta^{0.8451}\right )  \nonumber\\
&&+ 
       C_{7} \left (-0.0254 \zeta^{-0.3469} +
        0.0163 \zeta^{-0.1317} + 0.0336 \zeta^{0.4201} - 0.0246 \zeta^{0.8451}
\right ),
 \nonumber\\
	CC_5 & =& C_{4} \left (-0.2291 \zeta^{-0.3469} +  0.2167 \zeta^{-0.1317} + 
       0.0192 \zeta^{0.4201}  - 0.0067 \zeta^{0.8451}\right )  \nonumber\\
&&+ 
       C_{10} \left (-0.0095 \zeta^{-0.3469} -  0.0136 \zeta^{-0.1317} + 
       0.0264 \zeta^{0.4201}  - 0.0034 \zeta^{0.8451}\right ) \nonumber\\
&&  + 
      C_2 \left (-0.0413 \zeta^{-0.3469} + 0.0316 \zeta^{-0.1317} + 
         0.0120 \zeta^{0.4201}  - 0.0022 \zeta^{0.8451}
       \right )  \nonumber\\
&&+ C_{3} \left (-0.2102 \zeta^{-0.3469} +  0.2438 \zeta^{-0.1317} - 
       0.0336 \zeta^{0.4201} \right ) \nonumber\\
&& + 
      C_1 \left (-0.0319 \zeta^{-0.3469}   + 0.0452 \zeta^{-0.1317} - 
       0.0144 \zeta^{0.4201}  + 0.0011 \zeta^{0.8451}\right )  \nonumber\\
&&+ 
        C_{9} \left (0.0095 \zeta^{-0.3469}  + 0.0136 \zeta^{-0.1317} - 
       0.0264 \zeta^{0.4201}  +   0.0034 \zeta^{0.8451}\right ) \nonumber\\
&& + 
       C_5 \left (0.0380 \zeta^{-0.3469} +  0.9382 \zeta^{-0.1317}
 +         0.0078 \zeta^{0.4201} + 0.0159 \zeta^{0.8451}\right ) \nonumber\\
&&+ 
       C_{6} \left (-0.0240 \zeta^{-0.3469}   - 0.0305 \zeta^{-0.1317} + 
         0.0117 \zeta^{0.4201}  + 0.0427 \zeta^{0.8451}\right )  \nonumber\\
&&+ 
      C_8 \left (-0.0060 \zeta^{-0.3469}  - 0.0076 \zeta^{-0.1317} + 
       0.0029 \zeta^{0.4201}  + 0.0107 \zeta^{0.8451}  \right ) \nonumber\\
&&+ 
      C_{7} \left (0.0095 \zeta^{-0.3469}  +0.2346 \zeta^{-0.1317} \right .
 \nonumber\\
&&-\left.  
       0.25 \zeta^{-3/25} + 0.0020 \zeta^{0.4201} + 
       0.0040 \zeta^{0.8451} \right ),
 \nonumber\\
 	CC_6 &=& C_{4} \left (0.1825 \zeta^{-0.3469} -0.0784 \zeta^{-0.1317} + 
       0.0449 \zeta^{0.4201}  - 0.14894 \zeta^{0.8451}\right ) 
  \nonumber\\
   &&    + C_{10} \left (0.0075 \zeta^{-0.3469} + 0.0049 \zeta^{-0.1317} + 
        0.0617 \zeta^{0.4201}   - 0.07412 \zeta^{0.8451}
       \right ) \nonumber\\
&& + C_2 \left (0.0329 \zeta^{-0.3469} -0.0114 \zeta^{-0.1317} +
       0.0280 \zeta^{0.4201}  -  0.0495 \zeta^{0.8451}\right )  \nonumber\\
&&+ 
      C_{3} \left (0.1674 \zeta^{-0.3469}  -0.0882 \zeta^{-0.1317} - 
       0.0784 \zeta^{0.4201} -0.0007 \zeta^{0.8451}\right )  \nonumber\\
&&+ 
       C_1 \left (0.0254 \zeta^{-0.3469} - 0.0163 \zeta^{-0.1317} - 
       0.0336 \zeta^{0.4201}  +  0.0246 \zeta^{0.8451}\right ) \nonumber\\
&&  + 
      C_{9} \left (-0.0075 \zeta^{-0.3469}  - 0.0049 \zeta^{-0.1317} - 
       0.0617 \zeta^{0.4201}  + 0.07412 \zeta^{0.8451}
      \right )  \nonumber\\
&&+ C_5 \left (-0.0303 \zeta^{-0.3469}  - 0.3395 \zeta^{-0.1317} + 
       0.0182 \zeta^{0.4201} + 0.35157 \zeta^{0.8451}
       \right )  \nonumber\\
&&+ C_{6} \left (0.0191 \zeta^{-0.3469} +  0.0110 \zeta^{-0.1317} + 
       0.0274 \zeta^{0.4201} + 0.94253 \zeta^{0.8451}
       \right ) \nonumber\\
&&+ C_8 \left (0.0048 \zeta^{-0.3469} + 0.0028 \zeta^{-0.1317} + 
       0.0068 \zeta^{0.4201} \right . \nonumber\\
&&\left.~~ +0.2356 \zeta^{0.8451} - 0.25 \zeta^{24/25}\right ) + 
          C_{7} \left (-0.0076 \zeta^{-0.3469}  -  0.0849 \zeta^{-0.1317}
\right . \nonumber\\
&&\left. + 
       0.0833 \zeta^{-3/25} + 0.0046 \zeta^{0.4201} + 
       0.0879 \zeta^{0.8451} - 0.0833 \zeta^{24/25}\right ),
 \nonumber\\
 	CC_7&=& C_7 \zeta^{-3/25},
  \nonumber\\
 	CC_8 &=&C_{7} \left (-\zeta^{-3/25}+\zeta^{24/25}\right )/3. +C_8 
\zeta^{24/25} ,
  \nonumber\\
	CC_9&=& C_{10} \left (0.5 \zeta^{-6/25}-0.5 \zeta^{12/25}\right )+ 
          C_{9} \left ( 0.5 \zeta^{-6/25}+0.5 \zeta^{12/25} \right ), 
 \nonumber\\
	CC_{10} &=&C_{9} \left ( 0.5 \zeta^{-6/25}-0.5 \zeta^{12/25}\right ) + 
          C_{10} \left (0.5 \zeta^{-6/25} + 0.5 \zeta^{12/25}\right ), \label{w2}
 \end{eqnarray}
where $\zeta=\alpha_s(t)/\alpha_s(m_B)$. Here $\Lambda_{QCD}^{(4)}=250$MeV.

\end{appendix}


\vspace{1.0cm}
\begin{flushleft}
{\bf\Large Figure Captions}
\end{flushleft}
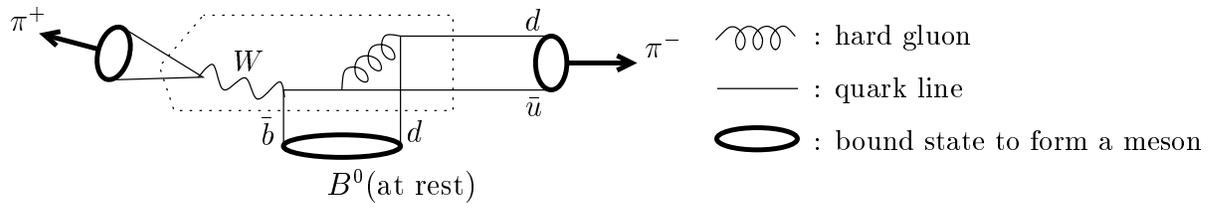
\begin{figure}[h]
 \caption{
 One of the decay process which contributes to $B\to\pi\pi$ decay. 
 $\bar{b}$ quark decays to produce a fast moving $\bar{u}$ quark.
    In general, this quark and $d$ quark are not lined up to form a
    pion. A gluon
    exchange is necessary in order that these quarks
    are lined up to form a pion. The part enclosed by dotted line
    describes the six-quark effective operator.}\label{fig:PhysicalPicB_1}
\end{figure}
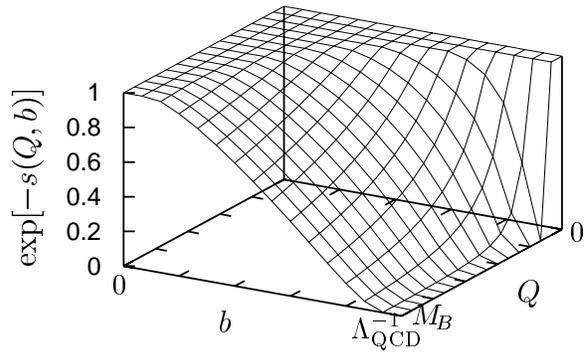
\begin{figure}[h]
 \caption{$b$-$Q$ dependence of $e^{-s}$. Nonperturbative
  region on $b\sim O(\Lambda^{-1})$ is suppressed by this exponent.
  Since the pion wave function has two Sudakov factor accompanied with 
  two light quarks, this suppression becomes much stronger.}\label{fig:sudakov}
\end{figure}
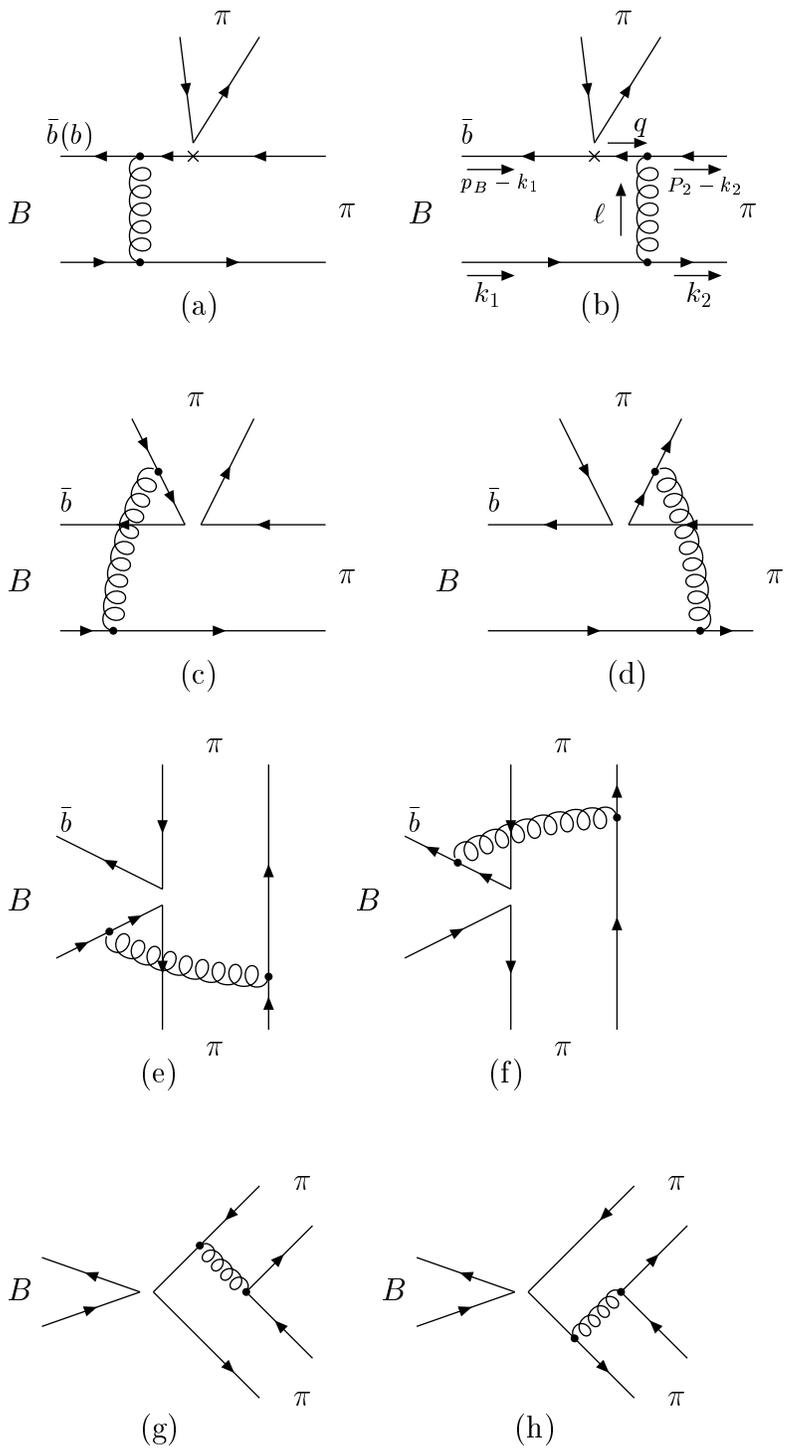
\begin{figure}[h]
 \caption{Diagrams contributing to the $B\to \pi\pi$ decays. The diagram
 (b) corresponds to Figure \ref{fig:PhysicalPicB_1}.}\label{fig:diagrams}
\end{figure}
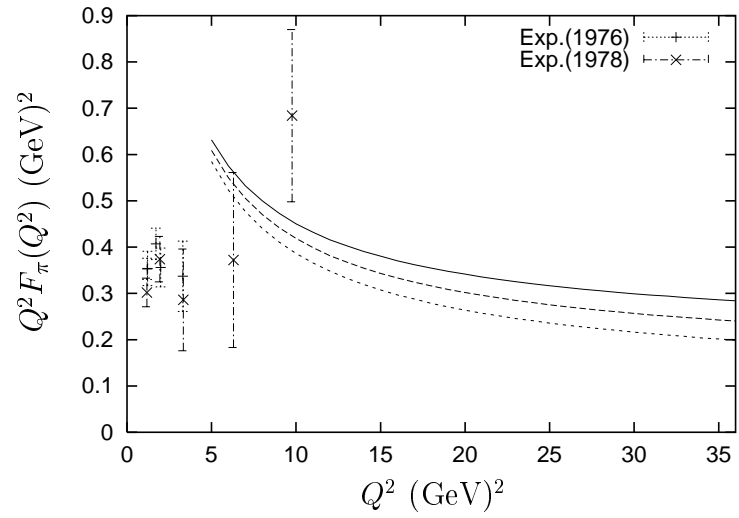
\begin{figure}[h]
\caption{$Q^2$ dependence for $F_\pi(Q^2)$ with the data \protect\cite{Bebek}. 
The solid, dashed, and dotted lines correspond to $a^A = 0.8, 0.4$, and
 $0$, respectively.}
\label{fig:Q2dep}
\end{figure}
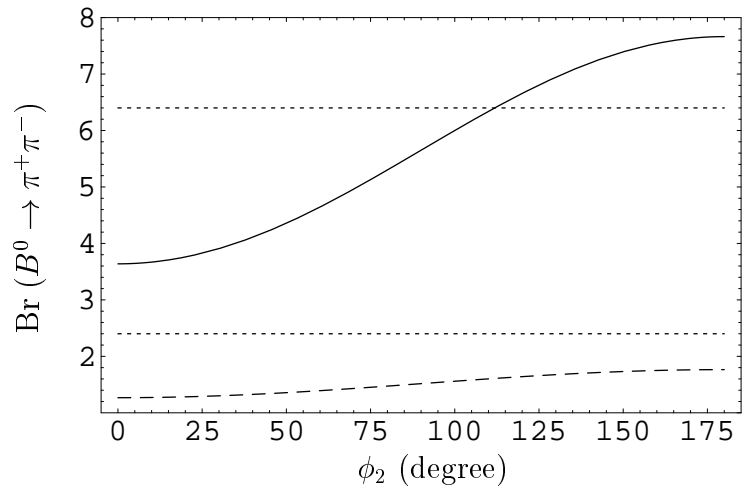
\begin{figure}[h]
 \caption{Averaged branching ratios (in unit of $10^{-6}$) 
 of $B^0(\bar B^0)\to \pi^+ \pi^-$, 
 $m_0=1.5$ GeV (solid line), $m_0=0$ GeV (dashed line).
The two dotted lines indicate the $1\sigma$ region of CLEO experiments in
eq.(\ref{cleodata}).}\label{bran}
\end{figure}
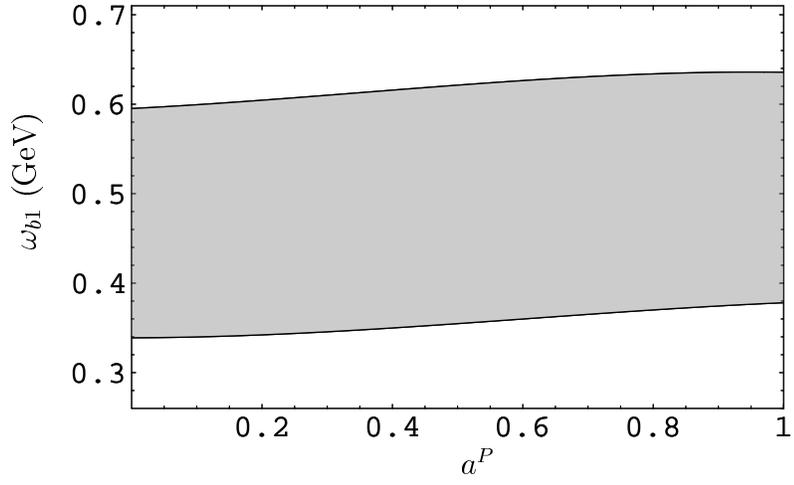
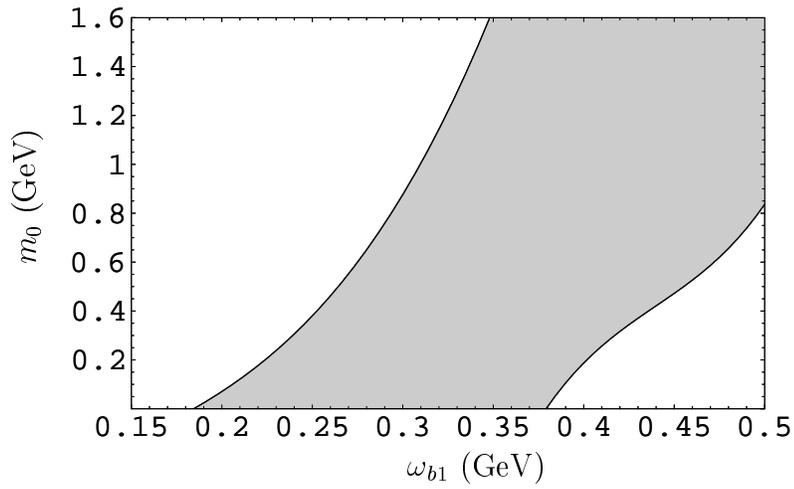
\begin{figure}[h]
\caption{Here we check the sensitivity of our calculation on parameter
$m_0$, $a^P$, and $\omega_{b1}$.
Others are defined in the begining of Section \ref{sc:NumCalc}.
The shaded areas are allowed by the data, eq.~(\ref{cleodata}), for
arbitrary $\phi_2$; a) we fix $m_0 = 1.5$ GeV and show the region for
 $a^P$ and $\omega_{b1}$; b) we fix $a^P = 0$ and show the allowed 
region for $\omega_{b1}$ and $m_0$. 
}\label{fig:region_pipi}
\end{figure}
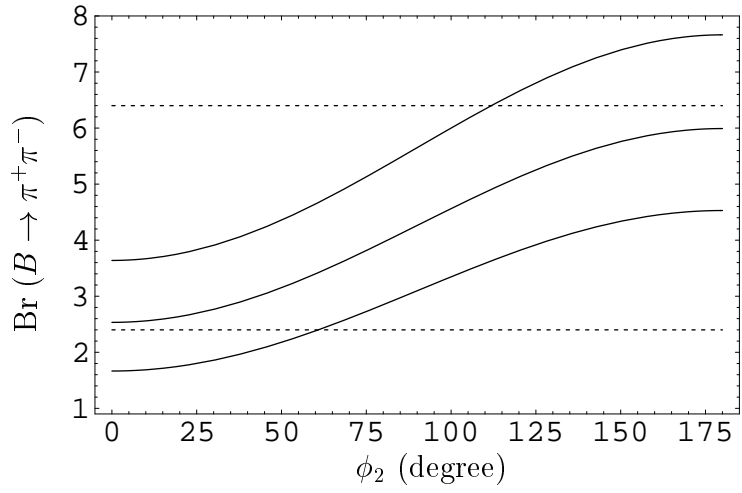
\begin{figure}[h]
\caption{Dependence on $a^A$ for the branching ratio (in unit of
 $10^{-6}$) of $B \to \pi^+\pi^-$.  $a^A = 0.8, 0.4, 0$ in descending
 order, respectively}\label{fig:varyCa}
\end{figure}
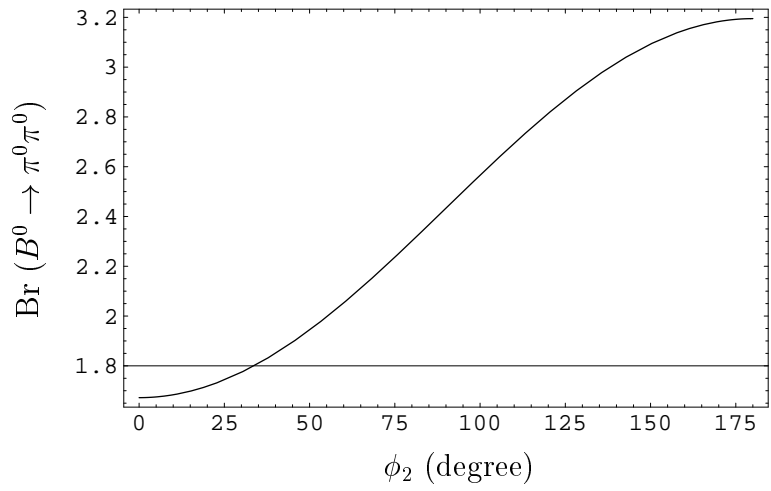
\begin{figure}[h]
\caption{Averaged branching ratios (in unit of $10^{-7}$)
of  $B^0\to \pi^0 \pi^0$
as a function
of CKM angle $\phi_2$.}\label{p0p0}
\end{figure}
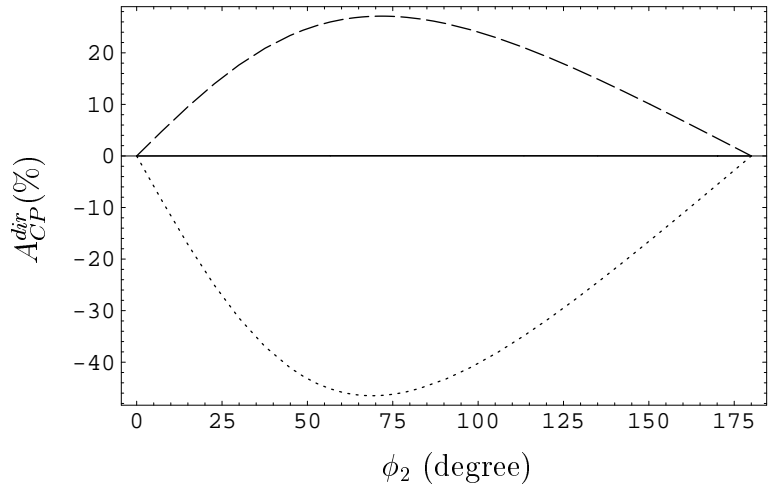
\begin{figure}[h]
\caption{Direct CP violation parameters (in percentage) 
of $B^0\to \pi^+ \pi^-$ (dotted line),
and $B^+\to \pi^+ \pi^0$ (solid line), and $B^0\to \pi^0 \pi^0$ 
(dashed line)
as a function
of CKM angle $\phi_2$.}\label{cppm}
\end{figure}
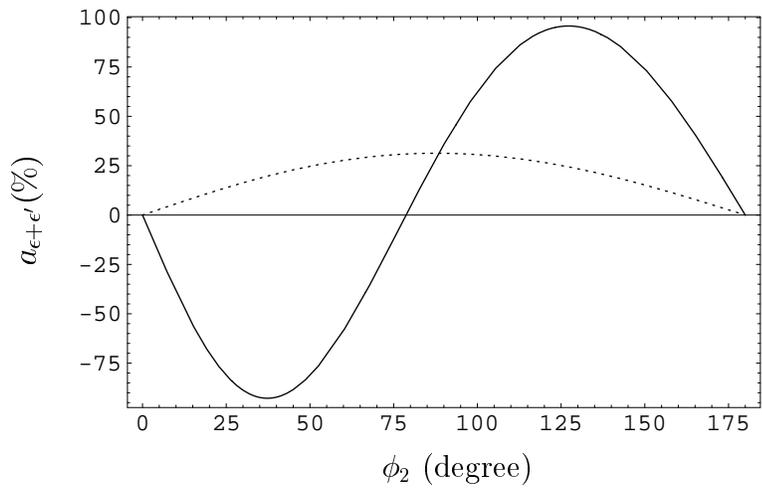
\begin{figure}[h]
\caption{CP violation parameters $a_{\epsilon+\epsilon'}$ (in percentage)
 of $B^0\to \pi^+ \pi^-$ (solid line),
and $B^0\to \pi^0 \pi^0$ (dotted line),
as a function
of CKM angle $\phi_2$.}\label{ae}
\end{figure}
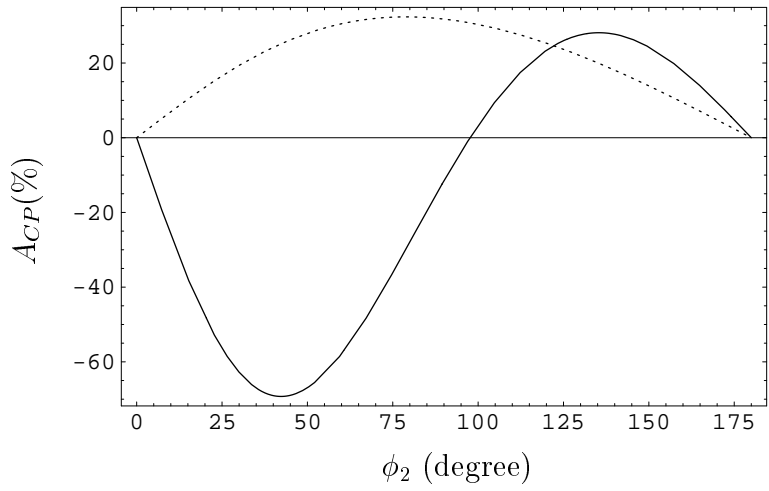
\begin{figure}[h]
\caption{The integrated CP asymmetries (in percentage)
 of $B^0\to \pi^+ \pi^-$ (solid line),
and $B^0\to \pi^0 \pi^0$ (dotted line),
as a function
of CKM angle $\phi_2$.}\label{acp}
\end{figure}

\end{document}